\documentclass[aps, prd, twocolumn, 
showpacs, showkeys, 
superscriptaddress, 
nofootinbib
]{revtex4-2}

\usepackage{physics}
\usepackage[T1]{fontenc}
\usepackage[utf8]{inputenc}
\usepackage[english]{babel}
\usepackage{microtype}

\usepackage{newtxtext}
\usepackage{newtxmath}

\usepackage{bm}
\usepackage{slashed}

\usepackage[dvipsnames,table,xcdraw]{xcolor}
\definecolor{lightyellow}{RGB}{255,250,205}

\usepackage{graphicx}
\usepackage{epstopdf}

\usepackage{array}        
\usepackage{tabularx}
\usepackage{multirow}
\usepackage{longtable}

\usepackage{url}
\usepackage{orcidlink}
\usepackage{subfigure}
\usepackage{soul}
\usepackage{xcolor}
\usepackage{physics}
\usepackage{booktabs}
\usepackage{ulem}

\usepackage{hyperref} 
\hypersetup{ 
  colorlinks=true,
  linkcolor=blue,  
  citecolor=cyan,
  urlcolor=blue,           
}

\begin{document}

\title{
In search for signals of the $D\bar{D}$ bound state  $X(3700)$ from study of the $B^+ \to D^+ D^- K^+$, $B^0 \to D^+ D^- K^0$ and $\Lambda_b \to D^+ D^- \Lambda$  reactions}

\author{Xiu-Lei Ren\orcidlink{0000-0002-5138-7415}}
\email[]{xiulei.ren@sdu.edu.cn}
\affiliation{Shandong Provincial Key Laboratory of Nuclear Science, Nuclear Energy Technology and Comprehensive Utilization, 
\& School of Nuclear Science, Energy and Power Engineering, Shandong University, 250061 Jinan, China }

\author{Hai-Peng Li\orcidlink{0009-0008-2985-3011}}
\email[]{haipeng@nnnu.edu.cn}
\affiliation{School of Physics and Electronics, Nanning Normal University, Nanning 530100, China}

\author{Wei-Hong Liang\orcidlink{0000-0001-5847-2498}}
\email[]{liangwh@gxnu.edu.cn}
\affiliation{Department of Physics, Guangxi Normal University, Guilin 541004, China}
\affiliation{Guangxi Key Laboratory of Nuclear Physics and Technology, Guangxi Normal University, Guilin 541004, China}

\author{Chu-Wen Xiao\orcidlink{0000-0001-5303-8350}}
\affiliation{Department of Physics, Guangxi Normal University, Guilin 541004, China}
\affiliation{Guangxi Key Laboratory of Nuclear Physics and Technology, Guangxi Normal University, Guilin 541004, China}

\author{Eulogio Oset\orcidlink{0000-0002-4462-7919}}
\email[]{Oset@ific.uv.es}
\affiliation{Department of Physics, Guangxi Normal University, Guilin 541004, China}
\affiliation{Departamento de F\'{\i}sica Te\'orica and IFIC, Centro Mixto Universidad de
		Valencia-CSIC Institutos de Investigaci\'on de Paterna, Apartado 22085,
		46071 Valencia, Spain}

\begin{abstract}
We perform a theoretical study of the $B^+ \to D^+ D^- K^+$, $B^0 \to D^+ D^- K^0$ and $\Lambda_b \to D^+ D^- \Lambda$ reactions by looking at the production mechanisms, with special emphasis in the final state interaction of the charmed mesons, which gives rise to the $X(3930)$, coupling strongly to $D_s\bar{D}_s$,  and another state that we call $X(3700)$, coupling strongly to $D \bar D$ that qualifies as a $D \bar D$ bound state. The combined study of all these reactions shows that the final state interaction responsible for the production of these resonances is more important in the $B^+ \to D^+ D^- K^+$ reaction than in the $\Lambda_b \to D^+ D^- \Lambda$ one. We have taken this into account and shown that normalizing the  $D^+ D^-$ mass distributions to the same value at the peak of the $\psi(3770)$ production, the two mass distributions are quite different in a range of 10 MeV above the  $D^+ D^-$ threshold, with a value about an order of magnitude bigger for the $B^+ \to D^+ D^- K^+$ reaction than for the $\Lambda_b \to D^+ D^- \Lambda$ one.
The $D^+D^-$ mass distribution by itself also shows an enhancement of about one order of magnitude in the presence of the $X(3700)$ with respect to a calculation where this resonance is absent.
We make a call to measure these magnitudes in the coming LHCb upgrades, which would bring great support to the existence of the bound $D^+ D^-$ state.
\end{abstract}


\date{\today}

\maketitle

\section{Introduction}
A $D\bar{D}$ bound state in isospin $I=0$ was predicted in Ref.~\cite{Gamermann:2006nm} using chiral dynamics, and has been posteriorly confirmed in other independent theoretical works~\cite{Nieves:2012tt,Hidalgo-Duque:2012rqv}. 
This state would be the analog of the $f_0(980)$, which stems from the interaction of $\pi \pi$ and $K \bar{K}$ originated by chiral dynamics \cite{Oller:1997ti,Oller:2000ma}, and can be considered as a $K\bar{K}$ bound state, something already suggested in Ref.~\cite{Weinstein:1990gu}. 
Since the strength of the interaction is bigger in analog states containing heavier quarks \cite{Ader:1981db,Zouzou:1986qh,Carlson:1987hh}, it is intuitive to expect that a corresponding bound $D\bar{D}$ state should also exist. 

Lattice QCD calculations in Ref.~\cite{Prelovsek:2020eiw} led to the conclusion that there were bound states for $D\bar{D}$ and $D_s \bar{D}_s$. 
A reanalysis of those lattice data done in Ref.~\cite{Shi:2024llv} reached the same conclusions. 
But further work along these lines has claimed a weaker attraction \cite{Wilson:2026bhu}. 
All these calculations are still done at relatively large pion masses, and the results depend somewhat on the coupled channels used in the analysis of the lattice data. 
The line of work is very useful and one expects to have more compelling results in the future. 

After the work of Ref.~\cite{Prelovsek:2020eiw} finding a $D_s \bar{D}_s$ bound state, the work of Ref.~\cite{Gamermann:2006nm} was reanalyzed in Ref.~\cite{Bayar:2022dqa} from the modern perspective given by the extension to the charm sector \cite{Molina:2010tx,Molina:2009ct} of the local hidden gauge approach~\cite{Bando:1984ej,Bando:1987br,Meissner:1987ge,Nagahiro:2008cv}. 
It was found that the dynamics generated simultaneously a state around $3700$ MeV, coupling mostly to $D\bar{D}$, and a state around $3930\, {\rm MeV}$, coupling mostly to $D_s \bar{D}_s$, which was identified with the $X(3930)$ seen in the $D^+D^-$ mass distribution in the $B^+ \to D^+D^-K^+$ decay~\cite{LHCb:2020pxc,LHCb:2020bls}. 
It was also shown that the state produces an enhancement in the $D^+_s D^-_s$ mass distribution close to threshold which is compatible with the LHCb observation in the $B^+\to D^+_sD^-_s K^+$ reaction \cite{LHCb:2022aki}. 
It is well known that a state lying below a hadron-hadron threshold and coupling strongly to that hadron-hadron component leads to a significant enhancement in the production cross section of this component in different reactions \cite{Dong:2021rpi}. 
Hence, there is a clear evidence for the $X(3930)$ state and its strong coupling to $D_s \bar{D}_s$. The coupled channel analysis of Ref.~\cite{Bayar:2022dqa} showed that the existence of the $D_s \bar{D}_s$ bound state had as a consequence the existence of a $D\bar{D}$ state.
  
There are other works that also predict a $D\bar{D}$ state using different phenomenological models \cite{Wong:2003xk,Zhang:2006ix,Liu:2009qhy,Li:2022shq,Wang:2026mfi}, and QCD sum rules also predict the existence of the two states within typical uncertainties of the method \cite{Mutuk:2022ckn}.  
  
Numerous reactions had been proposed to observe the $D\bar{D}$ state. 
In Ref.~\cite{Gamermann:2009ouq} the radiative decay of $\psi(3770)$ was suggested to observe this state, and a more elaborate description of the process was done in Ref.~\cite{Dai:2020yfu}. 
Three different methods were suggested in Ref.~\cite{Xiao:2012iq}, and looking for the decay of that state into $\gamma J/\psi$ was suggested in Ref.~\cite{Gamermann:2007bm}. 
In Ref.~\cite{Wei:2021usz} it was suggested to look for that state in the $\Lambda_b \rightarrow D\bar{D}\Lambda $ process. In Refs.~\cite{Xie:2022lyw,Wang:2026mfi} the investigation of the two states in $B^+ \to X K^+$ decays was also suggested. 
A different process  was proposed in Ref.~\cite{Brandao:2023vyg} to see the $D\bar{D}$ bound state, looking for the $B^+$ decay to $K^+ \eta \eta$, taking into account the decay of that state into light pseudoscalars which was investigated in Ref.~\cite{Xiao:2012iq}.
A similar strategy was also suggested in Ref.~\cite{Li:2023nsw}, but looking at the $\eta \eta_c$ decay. 
In Ref.~\cite{Shi:2021hzm} the production of the $D\bar{D}$ state in $pp$ and $p\bar{p}$ reactions was also proposed. 
The study with femtoscopic correlation functions has also been suggested as a method to see the $D\bar{D}$ bound state in Ref.~\cite{Abreu:2025jqy}.  

With so much study on these states and so many suggestions to see the elusive $D\bar{D}$ state, the experimental situation is still unclear. 
Some hints of the production of the state were observed in the $e^+ e^- \to J /\psi D \bar{D}, J /\psi D \bar{D}^*$ reactions in Ref.~\cite{Gamermann:2007mu}. 
The $e^+e^- \to J/\psi D\bar{D}$  reaction was further studied in Ref.~\cite{Wang:2019evy}, comparing to better data from Ref.~\cite{Belle:2017egg}, and the presence of the $D\bar{D}$ bound state was favored to interpret the data which observed a broad peak close to the $D \bar{D}$ threshold. 
Further support for the existence of the $D\bar{D}$ state was found in Ref.~\cite{Wang:2020elp} from the study of the $\gamma \gamma \to D \bar{D}$ reaction of the Belle \cite{Belle:2005rte} and Babar \cite{BaBar:2010jfn} collaborations, and a further study of that reaction done in Ref.~\cite{Deineka:2021aeu} claimed evidence for the $D\bar{D}$ bound state. 
  
With that situation, and the obvious interest that the issue has raised in the scientific community, we retake the problem and study simultaneously the $B^+ \to D^+ D^- K^+$, $B^0 \to D^+ D^- K^0$ and $\Lambda_b \to D^+ D^- \Lambda$ reactions, measured with relatively good precision by the LHCb collaboration in Refs.~\cite{LHCb:2020pxc,LHCb:2024hfo,LHCb:2019fns}, to make conclusions on the subject and suggest further specific experimental measurements that would allow more firm conclusions to be reached.  

\section{Formalism}
\subsection{$\Lambda_b \to D^+D^- \Lambda$ reaction}\label{sec:2A}
Our work follows closely the study of Ref.~\cite{Wei:2021usz}, but we use the $D\bar{D}$ and $D_s\bar{D}_s$ interactions from the work of Ref.~\cite{Bayar:2022dqa} based on the extension to the charm sector of the local hidden gauge approach~\cite{Bando:1984ej,Bando:1987br,Meissner:1987ge,Nagahiro:2008cv}, instead of the model of Ref.~\cite{Gamermann:2006nm} used in Ref.~\cite{Wei:2021usz}. 
The other modification is in the treatment of the $\psi(3770)$ excitation. After the work of Ref.~\cite{Wei:2021usz}, where predictions were made, there have been data for the reaction from the LHCb collaboration in Ref.~\cite{LHCb:2024hfo} and we can compare our result with these data. 

\begin{figure}[tbp]
    \centering
    \includegraphics[width=0.33\textwidth]{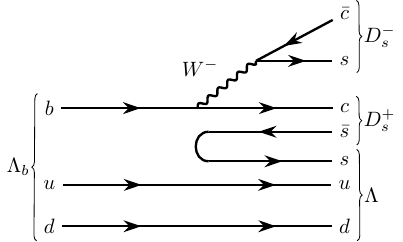}
    \caption{External emission and hadronization of the $cu$ (or $cd$) component.}
    \label{fig:1}
\end{figure}

\begin{figure}[tbp]
    \centering
    \includegraphics[width=0.33\textwidth]{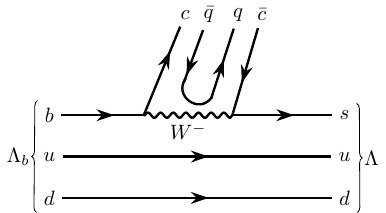}
    \caption{Internal emission and hadronization of the $c\bar{c}$ pair.}
    \label{fig:2}
\end{figure}

We describe below the $\Lambda_b \to D^+D^- \Lambda$ reaction at the quark level. 
To have the particle in the final state, we need to hadronize one pair of quarks, and we have two topologies contributing to the process, described in Figs.~\ref{fig:1} and \ref{fig:2}.  
One should keep in mind that the $\Lambda_b$ wave function is given by  \cite{Roberts:2007ni} 
$$
	\psi_{\Lambda_b} = \frac{1}{\sqrt{2}}\; b ( u d - du)\,\chi_{MA},
$$
where $\chi_{MA}$ denotes the mixed antisymmetric spin wave function and the $\Lambda$ hyperon has overlap with the $s(ud-du)\chi_{MA}$ component. 

After the hadronization of the $cu$ (or $cd$) quarks with an $\bar{s}s$ pair in external emission in Fig.~\ref{fig:1}, we obtain the component $D_s^- D_s^+\Lambda$, up to a global factor denoted by $A$, which is to be determined from the experimental data. In Fig.~\ref{fig:2}, we have internal emission and we hadronize the $c\bar{c}$ pair, which gives
\begin{align}\label{eq:1}
  c\bar{c} \to &~\sum\limits_i c\bar{q}_i q_i \bar{c} = \sum\limits_i P_{4i} P_{i4} = (P^2)_{44} \nonumber\\
  & = D^0\bar{D}^0 + D^+D^- +D_s^+D_s^-, 
\end{align}
where $P$ is the $q\bar{q}$ matrix written in terms of physical pseudoscalar mesons that we write below 
\begin{equation}
P=\begin{pmatrix}
\frac{\pi^0}{\sqrt{2}}+\frac{\eta}{\sqrt{3}}+\frac{\eta^{\prime}}{\sqrt{6}} & \pi^{+} & K^{+} & \bar{D}^0 \\[2mm]
\pi^{-} & -\frac{\pi^0}{\sqrt{2}}+\frac{\eta}{\sqrt{3}}+\frac{\eta^{\prime}}{\sqrt{6}} & K^0 & D^{-} \\[2mm]
K^{-} & \bar{K}^0 & -\frac{\eta}{\sqrt{3}}+\frac{2 \eta^{\prime}}{\sqrt{6}} & D_s^{-} \\[1.5mm]
D^0 & D^{+} & D_s^{+} & \eta_c
\end{pmatrix} ,
\label{eq:2}
\end{equation}
where the $\eta$-$\eta'$ mixing of Ref.~\cite{Bramon:1992kr} has been used. Neglecting terms with $\eta'$ and $\eta_c$ which do not play a role here, Eq.~\eqref{eq:1} and the other three quarks lead to the hadronized component 
\begin{equation}\label{eq:3}
	\frac{1}{\sqrt{2}}\;s(ud-du)\left(D^0\bar{D}^0 + D^+D^- + D_s^+D_s^-\right),
\end{equation}
and the $1/\sqrt{2}s(ud-du)$ three quark component has the same overlap with the physical $\Lambda$ state as with the mechanism of Fig.~\ref{fig:1}. 
However, internal emission is color suppressed \cite{Chau:1982da}, and then we give a weight $A\beta$ to this mechanism, having in mind that $\beta\simeq 1/3$. 

Altogether, the two mechanisms of  Figs.~\ref{fig:1} and \ref{fig:2} give rise after the weak decay to the hadronic components, 
\begin{equation}
	H = \left\{ A(1+\beta) D_s^+D_s^- + A\beta(D^+D^- + D^0\bar{D}^0) \right\} \Lambda.
\label{eq:4}
\end{equation}

\begin{figure*}[t]
    \centering
    \includegraphics[width=0.75\textwidth]{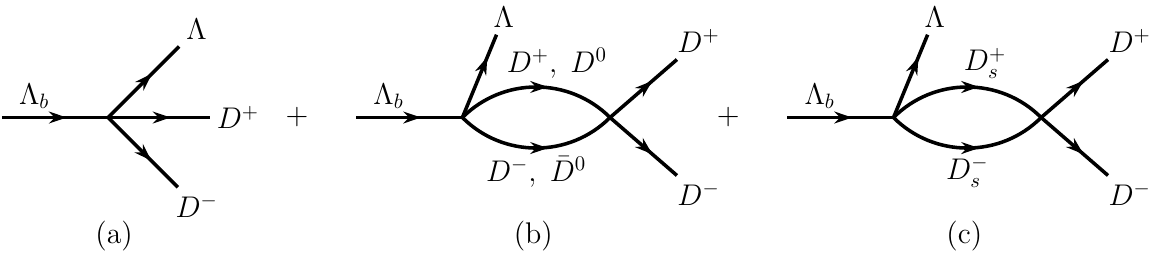}
    \vspace{-0.1cm}
    \caption{Terms contributing to the $\Lambda_b\to D^+D^-\Lambda$ decay: (a) tree level; (b) rescattering of $D^+D^-$ and $D^0\bar{D}^0$; (c) rescattering of $D_s^+D_s^-$.}
    \label{fig:3}
\end{figure*}

Our isospin phase convention for the multiplets is 
$(D^+,-D^0)$, $(\bar{D}^0, D^-)$. Hence the $I=0$ $D\bar{D}$ state is 
\begin{equation}
  |D\bar{D}, I=0\rangle = \frac{1}{\sqrt{2}} \left(|D^+D^-\rangle + |D^0\bar{D}^0\rangle \right),
\end{equation}
and we see that the contribution of states in $H$ of Eq.~\eqref{eq:4} has $I=0$. 
We should note that in $H$ we have the terms $D_s^+D_s^-\Lambda$ and $D^0\bar{D}^0\Lambda$ that do not correspond to the $D^+D^-\Lambda$ final state that we want to study. 
However, they can give contributions through final state interaction as shown in Fig.~\ref{fig:3}. 

Analytically, the terms of Fig.~\ref{fig:3} give rise to a transition $t$ matrix for $\Lambda_b \to D^+D^- \Lambda$, 
\begin{align}\label{Eq:tLamb_b}
	t_{\Lambda_b} & = A \beta + A \beta \,G_{D\bar{D}}(M_{D^+D^-}) \;t_{D\bar{D}, \, D\bar{D}} \nonumber\\
	 &\quad + \frac{1}{\sqrt{2}} \,A (1+\beta) \,G_{D_s \bar{D}_s}(M_{D^+D^-}) \;t_{D_s\bar{D}_s, \,D\bar{D}}, 
\end{align}
where $G_{D\bar{D}}$, $G_{D_s\bar{D}_s}$ are the loop functions of $D\bar{D}$ and $D_s\bar{D}_s$, respectively, appearing in the contribution of the $t$ matrix
$t_{D\bar{D},D\bar{D}}$, $t_{D_s\bar{D}_s, D\bar{D}}$ with the Bethe-Salpeter equation 
$
	T = \left[1-VG \right]^{-1} V 
$, 
with $V$ the transition potentials between the coupled channels. 
We follow strictly the steps of Ref.~\cite{Bayar:2022dqa} to calculate $G$ and $T$. 

\begin{figure}[t]
    \centering
    \includegraphics[width=0.28\textwidth]{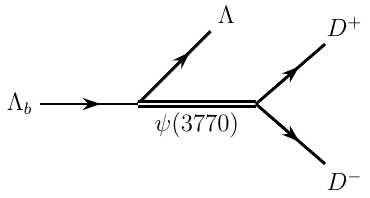}
    \caption{Diagram for $\Lambda_b\to \Lambda \psi(3770)$ and $\psi(3770) \to D^+D^-$ process.}
    \label{fig:4}
\end{figure}

\subsection{$\psi(3770)$ contribution} 
In the experiment of Ref.~\cite{LHCb:2024hfo}, a clear signal for $\psi(3770)$ excitation is seen in the $D^+D^-$ spectrum close to the $D^+D^-$ threshold. We depict this contribution in Fig.~\ref{fig:4}. 

The $\psi(3770)$ decays into $D^+D^-$ in $P$-wave, and the $\tilde{p}_{D^+}$ momentum is explicitly incorporated in Ref.~\cite{Wei:2021usz}. 
Here we include the full structure from the two vertices, although, given the small interference of this term with the others, the precise structure is not very important. 
The $\psi(3770)\to D^+D^-$ vertex gives us $\epsilon^\mu (p_{D^+}-p_{D^-})_\mu$ and the $\Lambda_b \to \Lambda\psi(3770)$ vertex that goes as $\gamma^\mu \epsilon_\mu$, for large mass of the baryons and up to a global normalization, can be parameterized as for the mesons, like $\epsilon^\nu(P_{\Lambda_b} + p_\Lambda)_\nu$. 
This form allows to get a simple formula in terms of the invariant masses shown in Ref.~\cite{Toledo:2020zxj}. 
Labelling the particles as $D^-(1)$, $D^+(2)$, $\Lambda(3)$,  we obtain an easy structure for the $\psi(3770)$ excitation as 
\begin{equation}
  \label{eq:7}
	t_\psi = A\,\gamma\;\frac{M_{23}^2-M_{13}^2}{M^2_{D\bar D}- M_\psi^2 + i M_\psi \Gamma_\psi} ,
\end{equation}
with $\gamma$ an unknown constant to be obtained from the data. 
To evaluate the $D^+D^-$ mass distribution $(M_{12})$, we use the PDG formula \cite{ParticleDataGroup:2024cfk} accommodated to the field normalization for baryons of Ref.~\cite{Mandl:2010book}, 
\begin{equation}
	\frac{\dd\Gamma}{\dd M_{12}\, \dd M_{23}} = \frac{1}{(2\pi)^3}\frac{1}{32 M_{\Lambda_b}^3} 2M_{12} 2M_{23} 2M_{\Lambda_b} 2 M_\Lambda \;\left|t^\mathrm{(tot)}_{\Lambda_b} \right|^2,
	\label{eq:8}
\end{equation}
where 
\begin{equation}
	t^\mathrm{(tot)}_{\Lambda_b} = t_{\Lambda_b} + t_\psi.
\end{equation}
In order to get $\dd\Gamma/\dd M_{12}$, we integrate Eq.~\eqref{eq:8} over $M_{23}$ with the limits of the PDG~\cite{ParticleDataGroup:2024cfk}. 

\begin{figure}[b]
    \centering
    \includegraphics[width=0.33\textwidth]{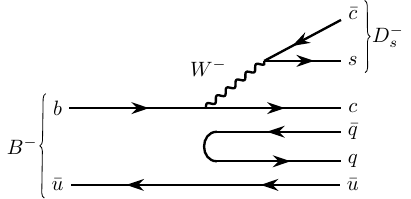}
    \caption{External emission in $B^-\to D_s^- c\bar u$ followed by $c\bar u$ hadronization.}
    \label{fig:5}
\end{figure}

\begin{figure}[t]
    \centering
    \includegraphics[width=0.33\textwidth]{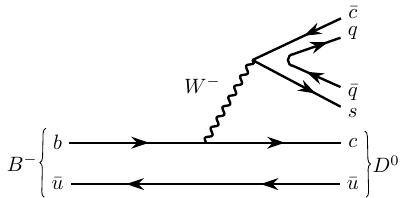}
    \caption{External emission for $B^-$ decay with hadronization of the $\bar cs$ pair.}
    \label{fig:6}
\end{figure}
\begin{figure}[t]
    \centering
    \includegraphics[width=0.33\textwidth]{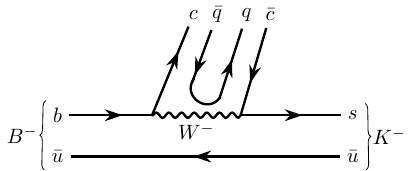}
    \caption{Mechanism of internal emission for $B^-$ decay with hadronization of the $c\bar c$ pair.}
    \label{fig:7}
\end{figure}

\subsection{$B^+\to D^+D^-K^+$ decay}
The reaction has been analyzed in the LHCb work of Ref.~\cite{LHCb:2020pxc}.
We proceed to describe it at the quark level as in the previous section~\ref{sec:2A}.
In Fig.~\ref{fig:5} we show the mechanism of external emission with hadronization of the $c\bar u$ pair.
This hadronization utilizes Eq.~\eqref{eq:2} to yield
\begin{equation}
  c\bar u\to\sum_i c\bar q_i q_i \bar u=\sum_i P_{4i}P_{i1}=(P^2)_{41},
  \label{eq:9}
\end{equation}
and hence we get the hadronic component
\begin{equation}
  \tilde{H}'=D_s^-\left[ D^0 \left( \frac{\pi^0}{\sqrt{2}}+\frac{\eta}{\sqrt{3}} \right)+D^+\pi^-+D_s^+K^- \right],
  \label{eq:10}
\end{equation}
with translated to the $B^+$ decay is 
\begin{equation}
  {H}'=D_s^+\left[ \bar D^0 \left( \frac{\pi^0}{\sqrt{2}}+\frac{\eta}{\sqrt{3}} \right)+D^-\pi^++D_s^-K^+ \right].
  \label{eq:11}
\end{equation}
Only the $D_s^+D_s^-K^+$ term can lead to the final state $D^+D^-K^+$ through final state interaction.

We can also have the hadronization of the $\bar cs$ pair in the external emission as shown in Fig.~\ref{fig:6}.
Following the same steps as before we have now 
\begin{equation}
  s\bar c\to \sum_i s\bar q_i q_i\bar c=\sum_i P_{3i}P_{i4}=\left(P^2\right)_{34},
  \label{eq:12}
\end{equation}
and we obtain  
\begin{equation}
      \tilde{H}''=D^0\left[ \bar D^0 K^- + D^- \bar K^0 -\frac{1}{\sqrt{3}} D_s^- \eta\right],
  \label{eq:13}
\end{equation}
with translated to $B^+$ decay gives as 
\begin{equation}
  H''=\bar D^0\left[D^0K^+ + D^+K^0 - \frac{1}{\sqrt{3}}D_s^+ \eta  \right].
  \label{eq:14}
\end{equation}
Here the only good component is $D^0 \bar D^0 K^+$, which upon final state interaction will produce $D^+D^-K^+$.

There is another way to reach the desired final state with the mechanism of internal emission, as shown in Fig.~\ref{fig:7}.
Once again we hadronize the $c\bar c$ pair leading to the $\left(P^2\right)_{44}$ combination, which translated to $B^+$ decay gives the hadronic combination
\begin{equation}
  H'''=K^+\left[ \bar D^0D^0+D^-D^++D_s^-D_s^+\right],
  \label{eq:141}
\end{equation}
and now the $D^+D^- K^+$ component contributes to tree level and all the terms through rescattering.
Then we give a weight $A'$ to the mechanism of Figs.~\ref{fig:5} and \ref{fig:6} with external emission and a weight $A'\beta$ to the mechanism of Fig.~\ref{fig:7} with internal emission
\footnote{As discussed in Ref.~\cite{Liang:2025jkj}, the mechanisms of Figs.~\ref{fig:5} and \ref{fig:6} do not have to have the same weight in principle, but a fit to data of different $B$ decays showed that the weights could be equal within errors. We take equal weights here.},
such that the final hadronic components relevant to in the $B^+$ decay are given by 
\begin{equation}
  H=K^+\left[A'\beta D^+D^-+A'(1+\beta)D^0\bar D^0+A'(1+\beta)D_s^+D_s^-\right].
  \label{eq:15}
\end{equation}

\subsection{$B^0\to D^+D^-K^0$ decay}
\begin{figure}[t]
    \centering
    \includegraphics[width=0.33\textwidth]{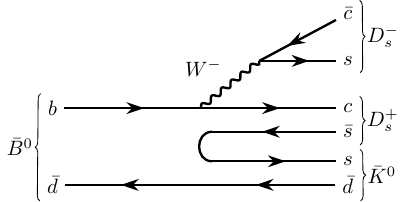}
    \caption{External emission for $\bar B^0$ decay with $c\bar d$ hadronization to $D_s^+\bar K^0$.}
    \label{fig:8}
\end{figure}

\begin{figure}[t]
    \centering
    \includegraphics[width=0.33\textwidth]{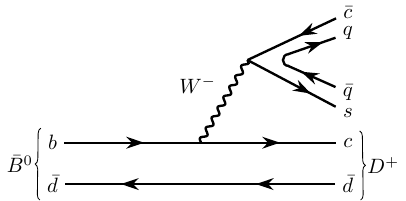}
    \caption{External emission for $\bar B^0$ decay with $\bar cs$ hadronization.}
    \label{fig:9}
\end{figure}

For $B^0\to D^+D^-K^0$ decay, the procedure is identical, with a small change in the particles produced.
We have first external emission, shown in Fig.~\ref{fig:8}, with $c\bar d$ for $\bar B^0$ decay.
This mechanism gives rise to $D_s^-D_s^+\bar K^0$ to which we give a weight. Since the number of counts in the experiment is much smaller than for $B^+$ decay, we must give these terms a different weight, which we call $A''$.
Next we have the mechanism of Fig.~\ref{fig:9}, also with external emission, to which we give a weight $A''$, as before.
The $\bar cs$ pair hadronizes as 
\begin{align}
  \nonumber
  s\bar c&\to\sum_i s\bar q_iq_i\bar c=\sum_i P_{3i}P_{i4}=\left(P^2\right)_{34}\\
         &\to D^+\left[K^-\bar D^0+\bar K^0D^--\frac{1}{\sqrt{3}}\eta D_s^-\right],
  \label{eq:16}
\end{align}
with translated to $B^0$ decay will be
\begin{equation}
  \bar H'=D^-\left[K^+D^0+K^0D^+-\frac{1}{\sqrt{3}}\eta D_s^+\right],
  \label{eq:17}
\end{equation}
and only the $D^+D^- K^0$ term will be operative.


\begin{figure}[t]
    \centering
    \includegraphics[width=0.33\textwidth]{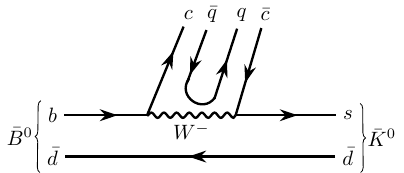}
    \caption{Mechanism for $\bar B^0$ decay from internal emission and $c\bar c$ hadronization.}
    \label{fig:10}
\end{figure}

Then we have the internal emission shown in Fig.~\ref{fig:10}, which hadronizes the $c \bar{c}$ pair as
\begin{align}
  \nonumber
  c\bar c&\to \sum_i c\bar q_iq_i\bar c=\sum_i P_{4i}P_{i4}=\left(P^2\right)_{44}\\
         &\to \bar K^0\left[D^0\bar D^0+D^+D^-+D_s^+D_s^-\right],
\end{align}
with translated to $B^0$ decay gives
\begin{equation}
  \bar H''=K^0\left[D^0\bar D^0+D^+D^-+D_s^+D_s^-\right],
  \label{eq:19}
\end{equation}
to which we give a weight $A''\beta$.
The sum of contributions for $B^0$ from the three mechanisms of Figs.~\ref{fig:8},~\ref{fig:9},~\ref{fig:10} gives
\begin{equation}
  \bar H=K^0\left[A''(1+\beta)D^+D^-+A''\beta D^0\bar D^0+A''(1+\beta)D_s^+D_s^-\right].
  \label{eq:20}
\end{equation}

\begin{figure}[t]
    \centering
    \includegraphics[width=0.28\textwidth]{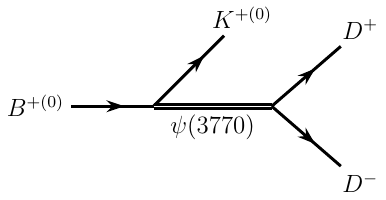}
    \caption{Direct excitation of the $\psi(3770)$ state in the $B^{+(0)}$ decay.}
    \label{fig:11}
\end{figure}
\begin{figure*}[htbp]
    \centering
    \includegraphics[width=0.75\textwidth]{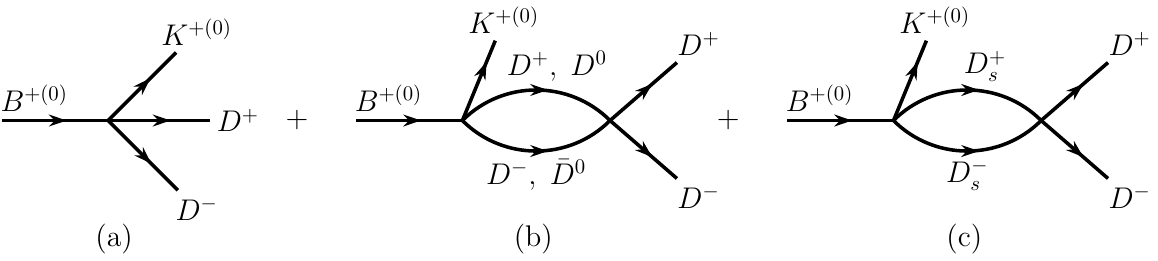}
    \caption{Diagrams for $B^{+}$ decay to $D^+D^-K^{+(0)}$: (a) tree level; (b) rescattering of $D^+D^-$ and $D^0\bar{D}^0$; (c) rescattering of $D_s^+D_s^-$.}
    \label{fig:12}
\end{figure*}

 \begin{figure*}[htbp]
  \includegraphics[width=0.40\textwidth]{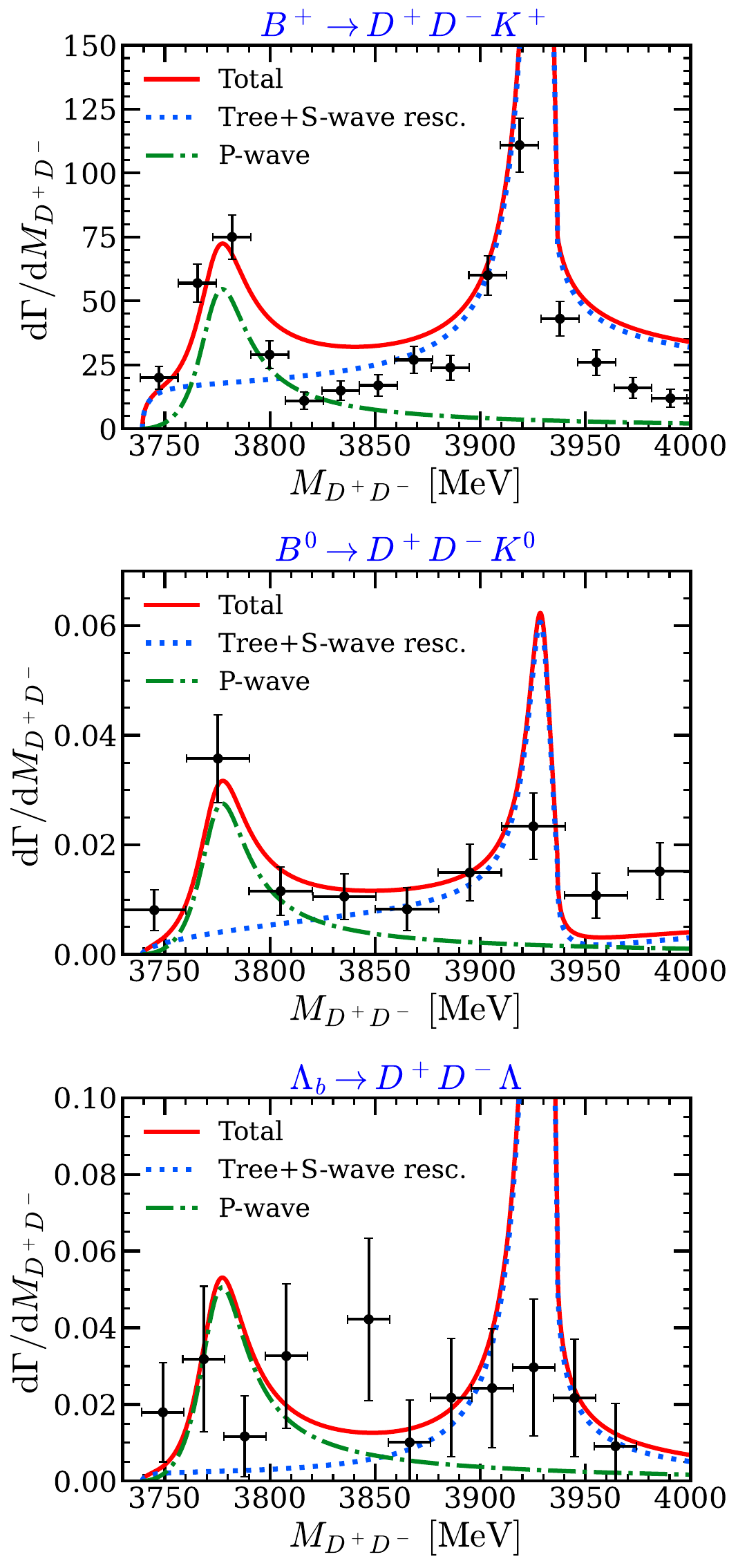}
  \includegraphics[width=0.40\textwidth]{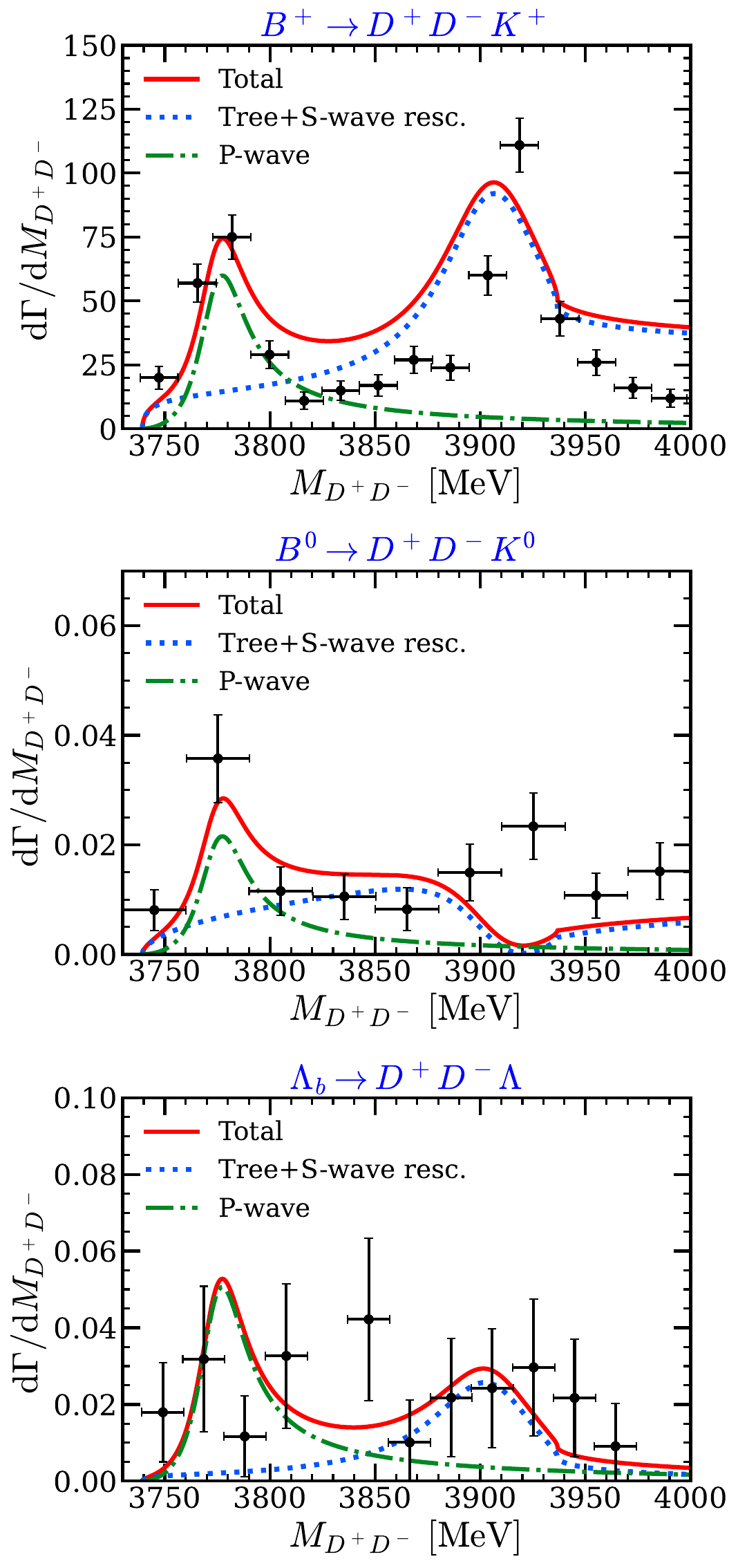}
  \caption{Combined fit results of the $D^+ D^-$ invariant mass distribution for the $B^+\to D^+D^- K^+$, $B^0\to D^+D^-K^0$ and $\Lambda_b\to D^+D^-\Lambda$ reactions, with left column and right column corresponding to Scenario I and Scenario II respectively (See Appendix \ref{APP} for details on Scenario I and Scenario II). 
  The experimental data are taken from LHCb measurements~\cite{LHCb:2020pxc,LHCb:2024hfo,LHCb:2019fns}. The solid lines denote our theoretical results, and the dotted and dot-dashed lines are the tree+$S$-wave rescattering and $P$-wave ($\psi (3770)$) contributions, respectively. 
  We fit only the data in the region from threshold up to $M_{D^+ D^-} \simeq 3800 \,{\rm MeV}$. The rest of the distributions are predictions.}
\label{Fig:x}
\end{figure*}

In addition, we will also have the contribution of the explicit excitation of the $\psi(3770)$ which is depicted in Fig.~\ref{fig:11}.
In this case, the vertices are $\epsilon^\mu(p_{D^+}-p_{D^-})_\mu$ and $\epsilon^\nu(P_B+p_K)_\nu$, and the contribution of the term is given by $t_\psi'$ and $t_\psi''$ for $B^+$ and $B^0$ decay respectively, with 
\begin{equation}
\begin{aligned}
  t'_\psi & =\frac{A'\gamma\;(M_{23}^2-M_{13}^2)}{M^2_{D\bar D}-M_\psi^2+iM_\psi\Gamma_\psi}, \\[1.5mm]
    t''_\psi & =\frac{A''\gamma\;(M_{23}^2-M_{13}^2)}{M^2_{D\bar D}-M_\psi^2+iM_\psi\Gamma_\psi}.
 \end{aligned}
  \label{eq:21}
\end{equation}
Since the $\psi(3770)$ has isospin $I=0$, the $BK\psi(3770)$ vertex is the same for $B^+\to K^+\psi$ and $B^0\to K^0\psi$.

Next we must proceed to implement final state interaction, which is shown diagrammatically in Fig.~\ref{fig:12} for $B^+$ and $B^0$ decays, respectively.
Analytically the diagrams in Fig.~\ref{fig:12} translate into 
\begin{align}\label{eq:221}
 t_{B^+} &=A'\beta+\frac{1}{2}\,A'(1+2\beta)\;G_{D\bar D}\;t_{D\bar D, \,D\bar D} \nonumber\\
 &\quad +\frac{1}{\sqrt{2}}\,A'(1+\beta)\;G_{D_s\bar D_s} \;t_{D_s \bar D_s, \,D \bar D}	, 
\end{align}
\begin{align}\label{eq:222}
	t_{B^0}& =A''(1+\beta)+\frac{1}{2} \,A''(1+2\beta) \;G_{D\bar D}\; t_{D\bar D, \,D\bar D} \nonumber\\
	&\quad +\frac{1}{\sqrt{2}}\, A''(1+\beta) \;G_{D_s\bar D_s} \;t_{D_s \bar D_s, \,D \bar D}.
\end{align}

We can see that these amplitudes are the same, up to the tree-level contribution.
The total amplitudes will be 
\begin{equation}
\begin{aligned}
  t_{B^+}^\text{(tot)}=t_{B^+}+t'_{\psi},\\[1.5mm]
  t_{B^0}^\text{(tot)}=t_{B^0}+t''_{\psi},
\end{aligned}
\end{equation}
and the double differential width is then given by (labeling $D^-$(1), $D^+$(2), and $K^{+(0)}$(3))
\begin{equation}
  \frac{\dd\Gamma}{\dd M_{12} \,\dd M_{23}}=\frac{1}{(2\pi)^3}\;\frac{1}{32m_B^3}\; 2M_{12}\, 2M_{23}\;\left|t^\mathrm{(tot)}_{B^{+(0)}}\right|^2.
  \label{eq:24}
\end{equation}
To obtain $\dd\Gamma/\dd M_{12}$, we integrate Eq.~\eqref{eq:24} over $M_{23}$ within the limits of the PDG~\cite{ParticleDataGroup:2024cfk}.

\begin{figure}[t]
  \includegraphics[width=0.415\textwidth]{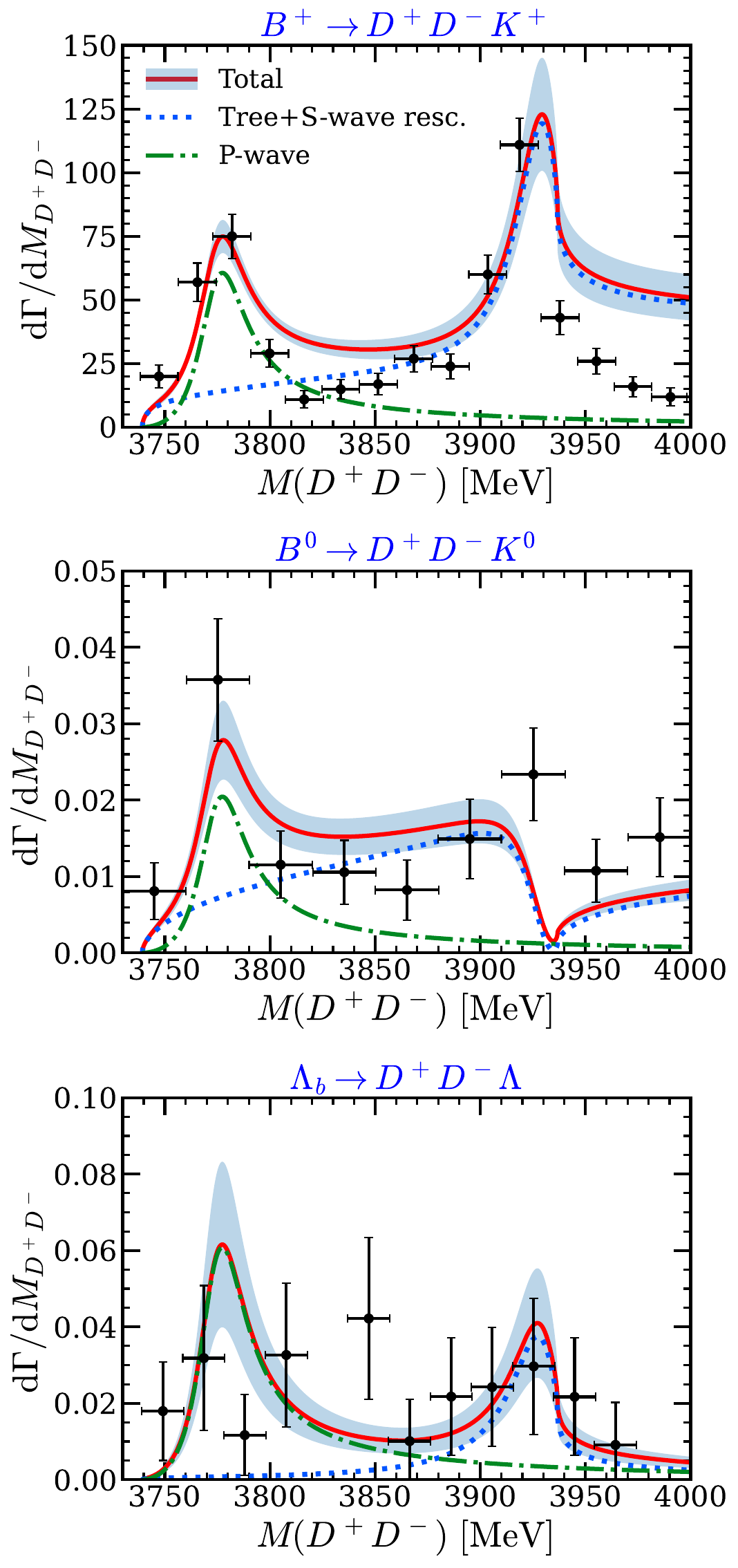}
  \vspace{-0.35cm}
  \caption{  
  Combined fit results of the $D^+ D^-$ invariant mass distributions for the $B^+\to D^+D^- K^+$, $B^0\to D^+D^-K^0$ and $\Lambda_b\to D^+D^-\Lambda$ reactions, corresponding to Scenario III (See Appendix \ref{APP} for details on Scenario III). 
  The error bands represent the uncertainties of the theoretical results, obtained by running with different random values of the parameters within their errors, with the constrains of the covariance matrix of Eq.~\eqref{eq:covariance}.  
  The experimental data are taken from LHCb measurements~\cite{LHCb:2020pxc,LHCb:2024hfo}. The solid lines denote our theoretical results, and the dotted and dot-dashed lines are the tree+$S$-wave rescattering and $P$-wave ($\psi (3770)$) contributions, respectively.
  We fit only the data in the region from threshold up to $M_{D^+ D^-} \simeq 3800 \,{\rm MeV}$. The rest of the distributions are predictions.}
  \label{Fig:comFit}
\end{figure}

To calculate the decay amplitudes $t_{\Lambda_b}, t_{B^+}, t_{B^0}$ of Eqs.~\eqref{Eq:tLamb_b}, \eqref{eq:221} and \eqref{eq:222}, we need the two-body scattering amplitudes $t_{D\bar D, D\bar D}$ and $t_{D_s \bar D_s, D\bar D}$, which are evaluated in Appendix~\ref{APP} with three different scenarios for the dynamical generation of the $X(3700)$ and $X(3930)$ states: Scenario I, Scenario II, and Scenario III.
They correspond to having a width or not of the $X(3700)$ state, and to changes in the position of the $X(3930)$ state. We will show the results for all three scenarios.

\section{Results}
We take the data for $B^+ \to D^+D^- K^+$ from Ref.~\cite{LHCb:2020pxc} and for $B^0 \to D^+D^-K^0$ and  $\Lambda_b\to D^+D^-\Lambda$  from Ref.~\cite{LHCb:2024hfo}. 
Due to different normalizations of the data, we take the parameters $A$ different for $\Lambda_b$, $B^+$, or $B^0$ decay. 
The parameter $\gamma$ is taken with the same value for all $\Lambda_b$, $B^+$ and $B^0$ decays. 
Similarly, the parameter $\beta$, which should be of the order of $1/N_c$, with $N_c$ the number of colors, is taken $\frac{1}{3}$ in all cases.
Later, we shall make small changes in $\beta$ to see the stability of the results.   

We make a combined fit to the data of the $D^+D^-$ mass distributions of all $B^+$, $B^0$, and $\Lambda_b$ decays, restricting ourselves to the region from threshold up to $M_{D^+ D^-} \simeq 3800\, {\rm MeV}$. 
This is a restricted region but in this way we avoid having to deal with potential contributions from the $\chi_{c2}(3930)$ as claimed in Ref.~\cite{LHCb:2020pxc}. 
Hence, we do not aim for a perfect description of the $X(3930)$ peak, and a qualitative agreement could be considered sufficient. 
The reason is that we do not pretend to obtain a good fit to all the data, but rather to investigate how, with improved statistics of these reactions, we can learn about the existence of the $D\bar{D}$ bound state. 

We present three sets of results corresponding to the three scenarios of Appendix~\ref{APP} mentioned above. 
In Fig.~\ref{Fig:x}, the left panel shows results obtained with two coupled channels, $D\bar{D}$ and $D_s\bar{D}_s$, using the regularization parameters in Eq.~\eqref{Eq:ScenarioI} for Scenario I, hence omitting the contribution from the $\eta\eta$ channel. 
The $\eta\eta$ channel was introduced empirically to account for the width of the predicted $X(3700)$ state, as found in Ref.~\cite{Xiao:2012iq}. 
The right panel of Fig.~\ref{Fig:x} displays the same mass distributions when three coupled channels are considered, using the regularization parameters in Eq.~\eqref{Eq:ScenarioII} for Scenario II. 
By comparing these results for Scenario I and Scenario II, we can see that the result for the first peak, corresponding to the $\psi(3770)$ excitation, does not change much, but the strength and even the shape of the second peak, corresponding to the $X(3930)$, do change, although a signal of that resonance is predicted in all the cases. 
It is interesting to see that by fitting only the region of the $\psi(3770)$ excitation, where the $X(3700)$ state below the $D^+D^-$ threshold does not show up, we can still obtain a qualitative signal for the $X(3930)$ excitation. 
The reason is the presence of the tree-level contribution, which is tied to the strength of both the $X(3700)$ excitation and the $X(3930)$ excitation, and this term gives some contribution in the region of the $\psi(3770)$ excitation. 

Furthermore, the position of the second peak is slightly lower than $3930\, {\rm MeV}$ in the right column of Fig.~\ref{Fig:x}. 
This is originated from the predicted mass and width of $X(3930)$ in the three coupled channels $D\bar{D}$, $D_s\bar{D}_s$, and $\eta\eta$, which are $3903.6\, {\rm MeV}$ and $82.3$ MeV, as given for Scenario II in Table~\ref{tab:2and3channels} of Appendix~\ref{APP}. 
By adjusting the subtraction constant $a_{D_s\bar{D}_s}$ as given in 
Eq.~\eqref{Eq:ScenarioIII} for Scenario III, 
we obtain a correct position of $X(3930)$ in Table~\ref{tab:2and3channels} of Appendix~\ref{APP}. 
Thus, we prefer to use the two-body amplitudes $t_{i, D\bar D}$ ($i=D\bar D, D_s\bar{D}_s$) corresponding to Scenario III to calculate the $\Lambda_b$, $B^{+(0)}$ decays,
and our favored combined fit results are presented in Fig.~\ref{Fig:comFit}, 
giving a reduced chi-square of $\chi^2/\text{d.o.f.}=3.8$.
The corresponding fitted parameters are determined as follows:
\begin{equation}
\label{Eq:finalPara} 
\begin{aligned}
    A' &=2426.2\pm448.4,  & A'' &= 41.46\pm7.37,  \\
    A  &= 76.89\pm 24.86, & \gamma &= 0.095\pm0.03, \\
    \beta &= 1/3~\text{(fixed)}.&&
  \end{aligned}
\end{equation}
These parameters correspond to Eq.~\eqref{Eq:tLamb_b} ($A, \beta$), Eqs.~\eqref{eq:7} and \eqref{eq:21} ($\gamma$), Eq.~\eqref{eq:221} ($A'$), and Eq.~\eqref{eq:222} ($A''$).
We do not give dimensions to these parameters since since we compare to experimental counts.
The covariance matrix $C$ of the parameters $A'$, $A''$, $A$, $\gamma$ is given by
\begin{equation}\label{eq:covariance}
C= 
\begin{pmatrix}
 2.011 \times 10^5 & 2725.720 & 8756.160 & -12.743 \\
 2725.720 &   54.300 &  123.433 &  -0.180 \\
 8756.160 &  123.433 &  618.231 &  -0.577 \\
  -12.743 &   -0.180 &   -0.577 & 8.398 \times 10^{-4}
\end{pmatrix}.
\end{equation}
From the correlation coefficients 
\begin{equation}
\rho_{ij}=\dfrac{C_{ij}}{\sqrt{C_{ii}\,C_{jj}}},
\end{equation}
we can see that $A'$ and $\gamma$ are relatively correlated \big($\abs{\rho_{14}}>0.9$\big),
but the rest of the pairs of parameters are only moderately correlated \big($\abs{\rho_{ij}}\lesssim 0.7$\big).
The relatively large correlations between the parameters mostly indicate that there is a trade off in the contribution to $\chi^2$ from the data of the different reactions.
Also the big strength of the $\psi(3770)$ peak is tied to the product of $A_i\gamma \,(A_i=A',\ A'',\ A)$, and not to the value of the individual parameters.

Once we see the kind of agreement that we get, we come to the main purpose of the present work: what should be done to observe a signal of the $X(3700)$ state? 
What we expect is that the presence of this state below the $D^+D^-$ threshold should produce an enhancement of the mass distributions close to the threshold. 
Unfortunately, the data are not precise enough to see that, and the presence of the strong $\psi(3770)$ peak in all the three reactions, not far from threshold ($3739$ MeV), certainly does not help.
Yet, very close to threshold, we should see the effects of this state. 

\begin{figure}[b]
  \includegraphics[width=0.42\textwidth]{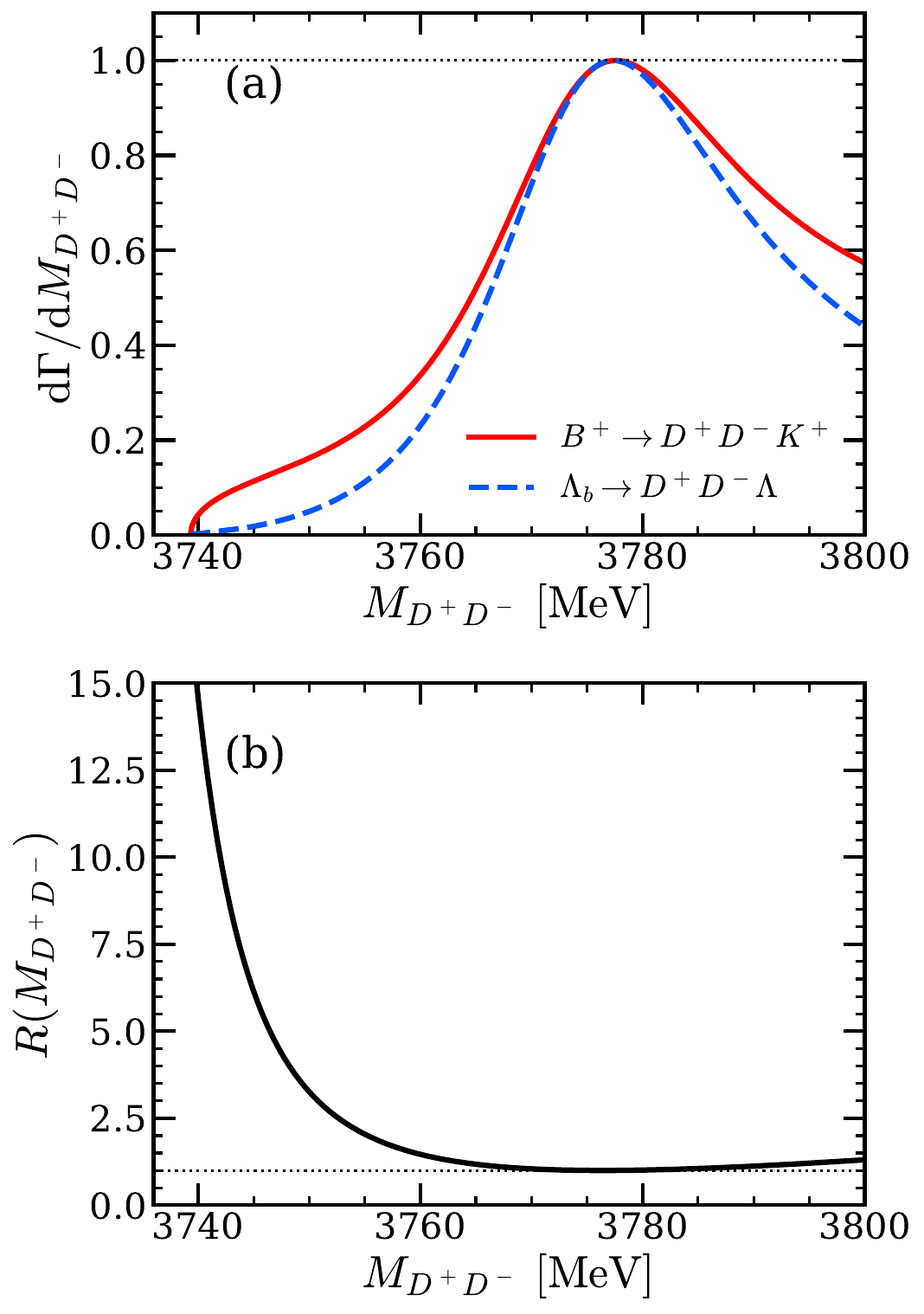}
  \vspace{-0.35cm}
  \caption{(a): $\dd\Gamma/\dd M_{D^+D^-}$  for the $\Lambda_b \to D^+D^-\Lambda$ and $B^+\to D^+D^-K^+$ reactions normalised to the same value at the peak of the $\psi(3770)$. (b): Ratio $R(M_{D^+ D^-})$ of Eqs.~\eqref{eq:ratioR} and \eqref{eq:ratioRtilde} for the two reactions as a function of $M_{D^+D^-}$, when the two differential widths have been normalised at the peak of the $\psi(3770)$ excitation.}
  \label{Fig:ratio}
\end{figure}

To understand the reasons of the proposal that we make here, we go back to Eq.~\eqref{Eq:tLamb_b} and see the term with $t_{D\bar{D},D\bar{D}}$ that would account for the contribution of the $X(3700)$ state goes with weight $A\beta$. 
On the other hand, $t_{B^+}$ from Eq.~\eqref{eq:221} has a weight $1/2A'(1+2\beta)$ in the $t_{D\bar{D},D\bar{D}}$ amplitude, and so has $t_{B^0}$ of Eq.~\eqref{eq:222} (with weight $A''$). 
But $t_{B^+}$ has a tree level that goes as $A'\beta$ while it has a weight $A''(1+\beta)$ in $t_{B^0}$. If we wish to see the effects of the $X(3700)$ resonance through $t_{D\bar{D},D\bar{D}}$, it is better to have a small tree level to magnify the relative effect of the resonance. 
This said, a comparative study of the $\Lambda_b \to D^+D^-\Lambda$ and $B^+ \to D^+D^-K^+$ reactions close to the threshold should give information concerning this point. 
Actually, one already hints at a bigger enhancement close to threshold for the $B^+\to D^+D^-K^+$ reaction than for the $\Lambda_b \to D^+D^-\Lambda$, as seen in Fig.~\ref{Fig:comFit}, but we show that comparison better in Fig.~\ref{Fig:ratio}. 

What we do in Fig.~\ref{Fig:ratio}(a) is to take the two mass distributions for $\Lambda_b \to D^+D^-\Lambda$ and $B^+\to D^+D^-K^+$ and normalize them to the same ratio at the peak of the $\psi(3770)$ excitation. 
We then examine the behavior very close to the threshold. We observe that the mass distribution for the $B^+\to D^+D^-K^+$ reaction shows a clear enhancement close to threshold, which is not present in the $\Lambda_b \to D^+D^-\Lambda$ reaction. 

It will be interesting to show the ratio of these two mass distributions, normalized to the peak of the $\psi(3770)$ as a function of the $D^+D^-$ invariant mass. 
We call this ratio $R(M_{D^+ D^-})$, given by
\begin{equation}\label{eq:ratioR}
  R(M_{D^+ D^-}) = \dfrac{\tilde{R} (M_{D^+ D^-})}{\tilde{R} (M_{D^+ D^-}=3770 \text{ MeV})},
\end{equation}
with 
\begin{equation}\label{eq:ratioRtilde}
  \tilde{R} (M_{D^+ D^-}) = \dfrac{\dd\Gamma/\dd M_{D^+D^-}(B^+\to D^+D^-K^+)}{\dd\Gamma/\dd M_{D^+D^-}( \Lambda_b \to D^+D^-\Lambda)}.
\end{equation}
In Fig.~\ref{Fig:ratio}(b), we show the ratio $R(M_{D^+ D^-})$ as a function of the $D^+D^-$ invariant mass, and we see that this ratio goes from $1$ at the peak of the $\psi(3770)$, by construction, to about $15$ when approaching the $D^+D^-$ threshold.
Such a huge factor is due to the presence of the $X(3700)$ state below the $D^+D^-$ threshold and should be seen in a devoted experiment.  

The implementation of this test should require precise measurements of $\dd \Gamma/\dd M_{D^+D^-}$ in the invariant mass region $M_{D^+D^-} \in [3739, 3750]$ MeV for both the $\Lambda_b \to D^+D^-\Lambda$ and $B^+\to D^+D^-K^+$ reactions. 
For the moment, there is only one data point in this region, and the errors are relatively large, particularly for the $\Lambda_b \to D^+D^-\Lambda$ reaction. 
However, the demanding task for this test can be accomplished in the upcoming upgrades of the LHCb facility. 
The detailed study done here should provide enough motivation to do it. 

\section{Model Dependence and Uncertainty Estimation}
\begin{figure}[htbp]
  \centering
  \includegraphics[width=0.48\textwidth]{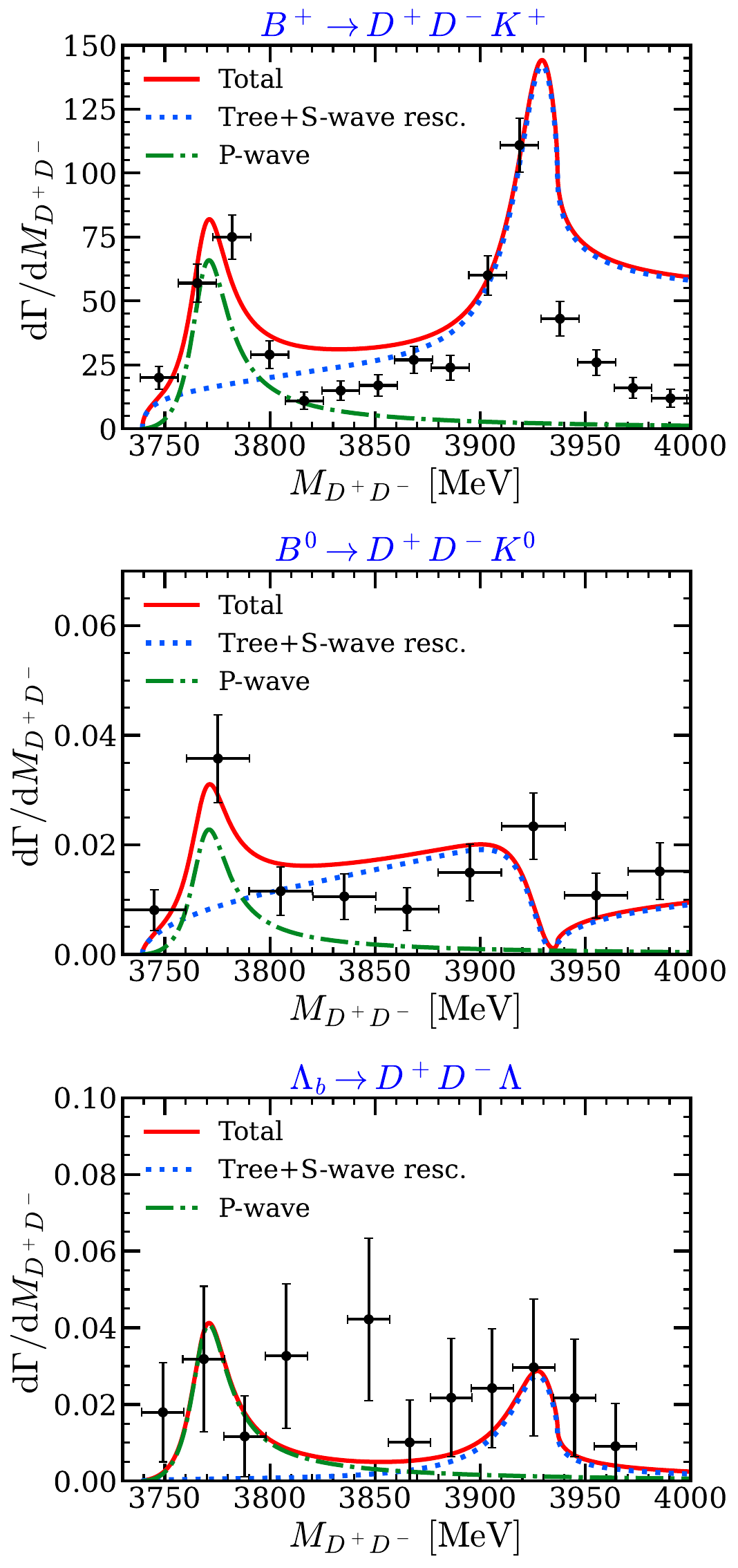}
  \caption{$D^+D^-$ invariant mass distribution, same as Fig.~\ref{Fig:comFit} but obtained using the energy-dependent width $\Gamma_\psi(M_{D^+D^-})$.}
  \label{fig:r1}
\end{figure}

\begin{figure}[htbp]
  \centering
  \includegraphics[width=0.48\textwidth]{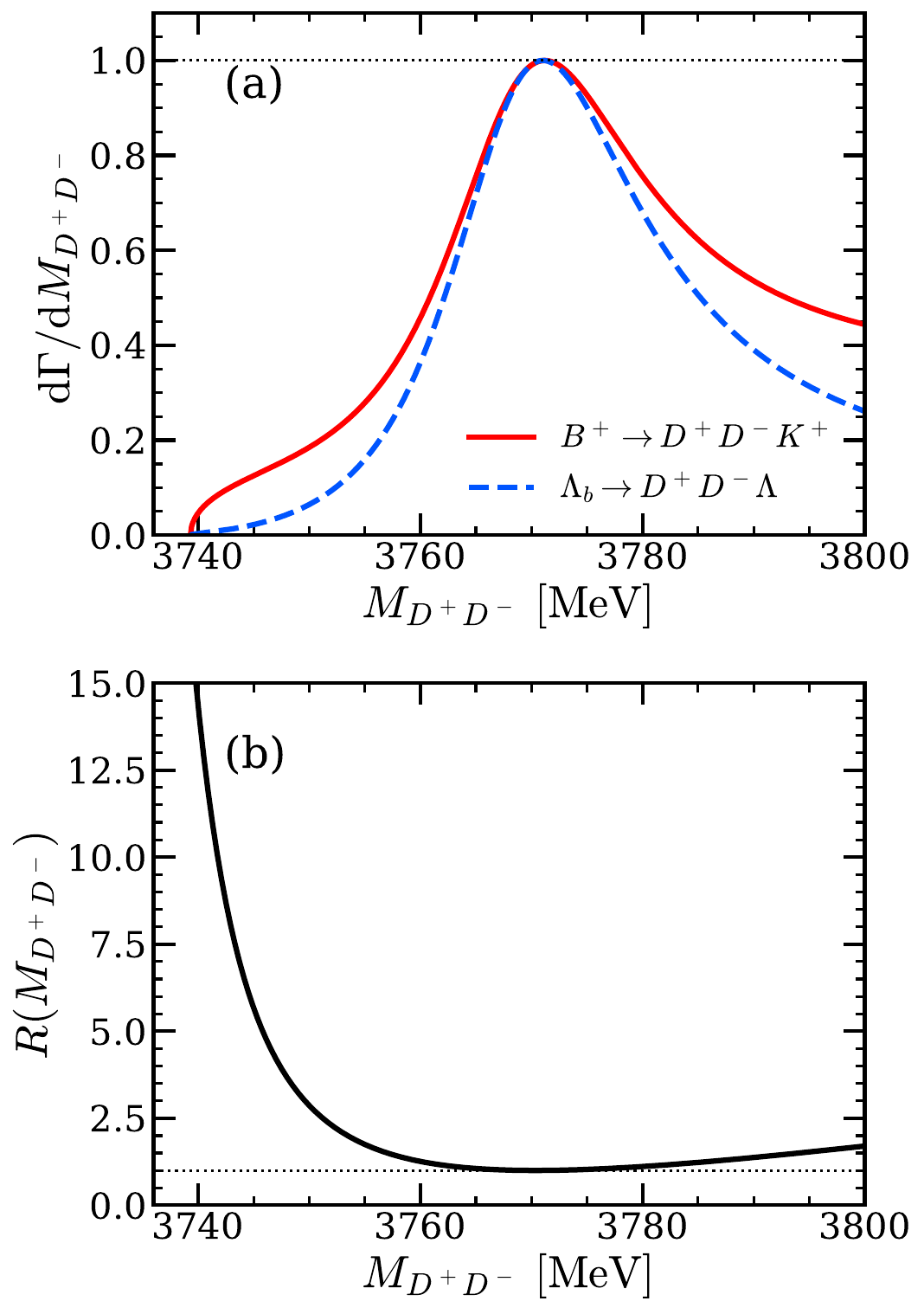}
  \caption{Same as Fig.~\ref{Fig:ratio}, but obtained using the energy-dependent width $\Gamma_\psi(M_{D^+D^-})$.}
  \label{fig:r2}
\end{figure}

In the amplitudes of the $\psi(3770)$ state given in Eqs.~\eqref{eq:7} and \eqref{eq:21}, we used a constant $\psi(3770)$ width, $\Gamma_\psi$.
Now, we replace the constant width with an energy-dependent one,
considering its decay to $D\bar D$ that practically accounts for the total width,
\begin{equation}
  \Gamma_\psi(M_{D\bar D}) = \Gamma_{\text{on}} \left( \frac{p_D(M_{D\bar D})}{p_D(M_\psi)} \right)^3,
\end{equation}
with
\begin{equation}
  p_D(M) = \frac{\lambda^{1/2}(M^2, m_D^2, m_D^2)}{2M},
\end{equation}
and $M_\psi$ and $\Gamma_\text{on}$ the nominal mass and width of the $\psi(3770)$.
With the modified amplitude, we refit the experimental data.
The new fit yields a reduced chi-square of $\chi^2/\text{d.o.f.}=2.8$, and the determined parameters are
\begin{equation}
\begin{aligned}
    A' &=2463.5\pm349.5,  & A'' &= 45.86\pm6.05,  \\
    A  &= 61.02\pm 19.41, & \gamma &= 0.099\pm0.02, \\
    \beta &= 1/3~\text{(fixed)}.&&
  \end{aligned}
\end{equation}
We observe an improvement of $\chi^2/\text{d.o.f.}$, and compared to Eq.~\eqref{Eq:finalPara} the parameters have only changed moderately.
The resulting invariant mass distributions and ratios are shown in Figs.~\ref{fig:r1} and \ref{fig:r2}, respectively, which are very similar to those in Figs.~\ref{Fig:comFit} and \ref{Fig:ratio}.
A direct comparison reveals that the fit quality and extracted parameters exhibit very small differences from those obtained using a constant width in Figs~\ref{Fig:comFit} and \ref{Fig:ratio}.


We take advantage here to see the sensitivity of the ratio $R(M_{D^+ D^-})$ to details of the model.
We have so far discussed three scenarios for the dynamical generation of the $X(3700)$ and $X(3900)$ states: Scenario I, Scenario II, and Scenario III (see Appendix~\ref{APP} for details).
By looking at Table~\ref{tab:2and3channels}, we see that the introduction of the $\eta\eta$ channel changed the mass of the $X(3930)$ state but barely changed the mass of the $X(3700)$ in $2\, {\rm MeV}$.
However, it was responsible for the appearance of a width in the $X(3700)$ state.
Comparing with Scenario II, Scenario III only changed the mass of the $X(3700)$ by $1\, {\rm MeV}$.
Our prefered scenario is Scenario III, which fits better the $X(3930)$ state, but we are mostly concerned about the $D^+D^-$ threshold behavior, to which we come back below. Before that we want to show the sensitivity of the resutls to changes of $\beta$ and the $\eta\eta$ transition potential $v_\eta$.

\begin{figure}[htbp]
  \centering
  \includegraphics[width=0.48\textwidth]{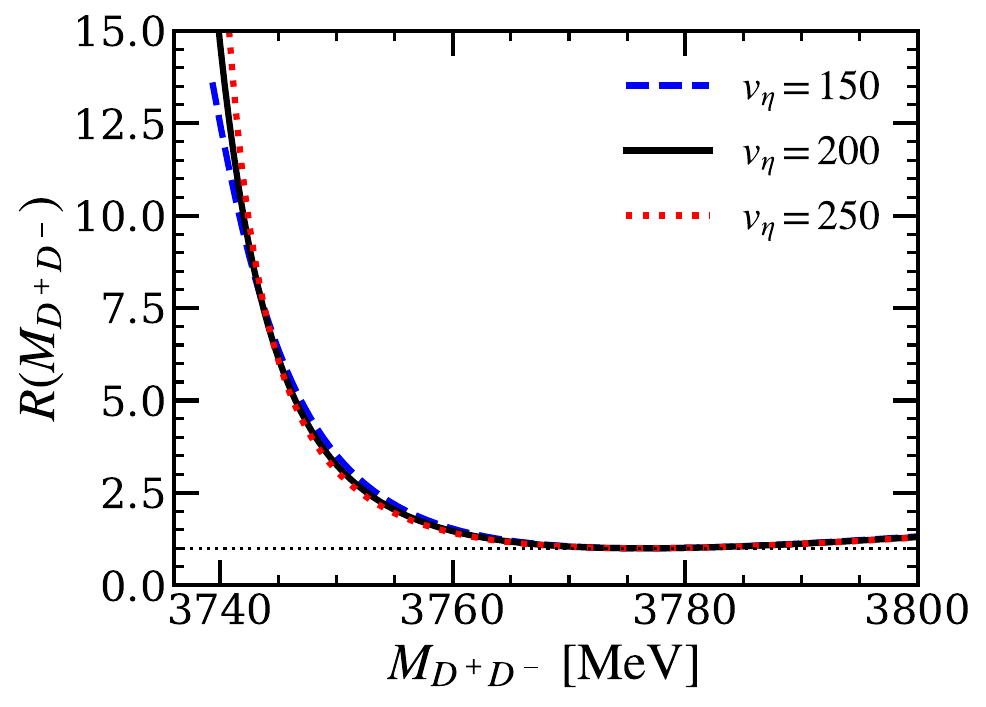}
  \includegraphics[width=0.48\textwidth]{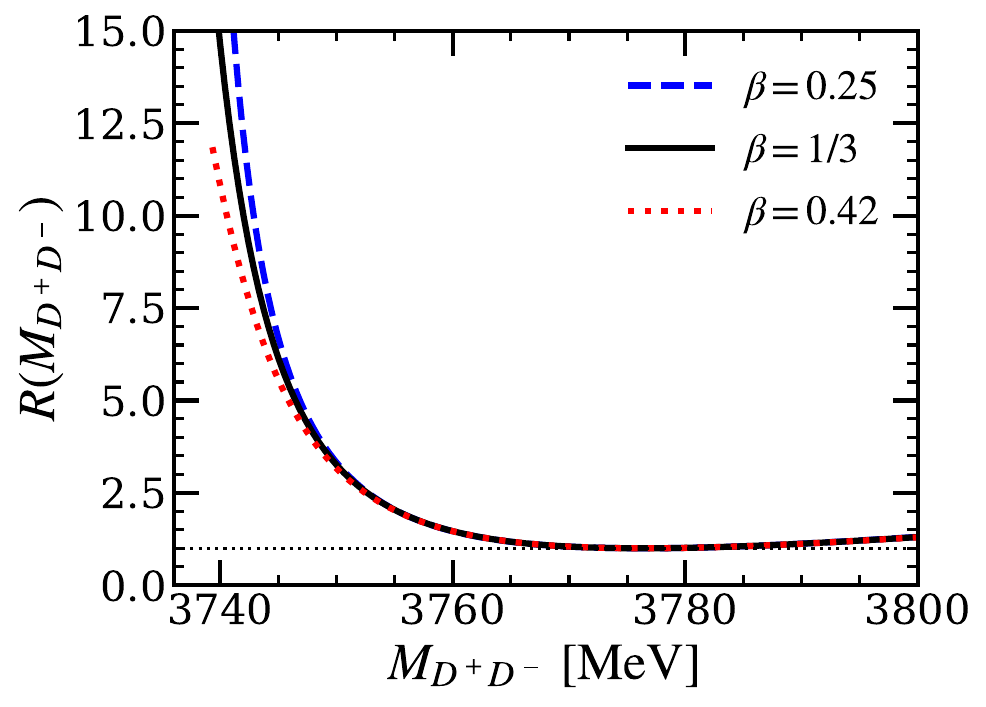}
  \caption{Sensitivity of the ratio $R(M_{D^+ D^-})$ to parameter variations. Panel (a) shows the ratios obtained by varying $v_\eta$ by $\pm 25\%$ from its original value ($v_\eta = 150, 200, 250$). Panel (b) illustrates the corresponding ratio under a $\pm 25\%$ variation of $\beta$ ($\beta = 0.25, 0.33, 0.42$). In all cases, the remaining parameters are refitted to the data.}
  \label{Fig:parameter_sensitivity}
\end{figure}

To examine the sensitivity of our predictions to the parameters $v_\eta$ and $\beta$, 
we vary their values by $\pm25\%$ around their original values and refit the experimental data.
Figure~\ref{Fig:parameter_sensitivity} displays the resulting ratio of $\dd\Gamma/\dd M_{D^+ D^-}$ for the $B^+$ and $\Lambda_b$ decays, in analogy to Fig.~\ref{Fig:ratio}(b).
As can be seen,  
the ratio exhibits only weak sensitivity to variations in both $v_\eta$ and $\beta$.

\begin{figure}[htbp]
  \centering
  \includegraphics[width=0.48\textwidth]{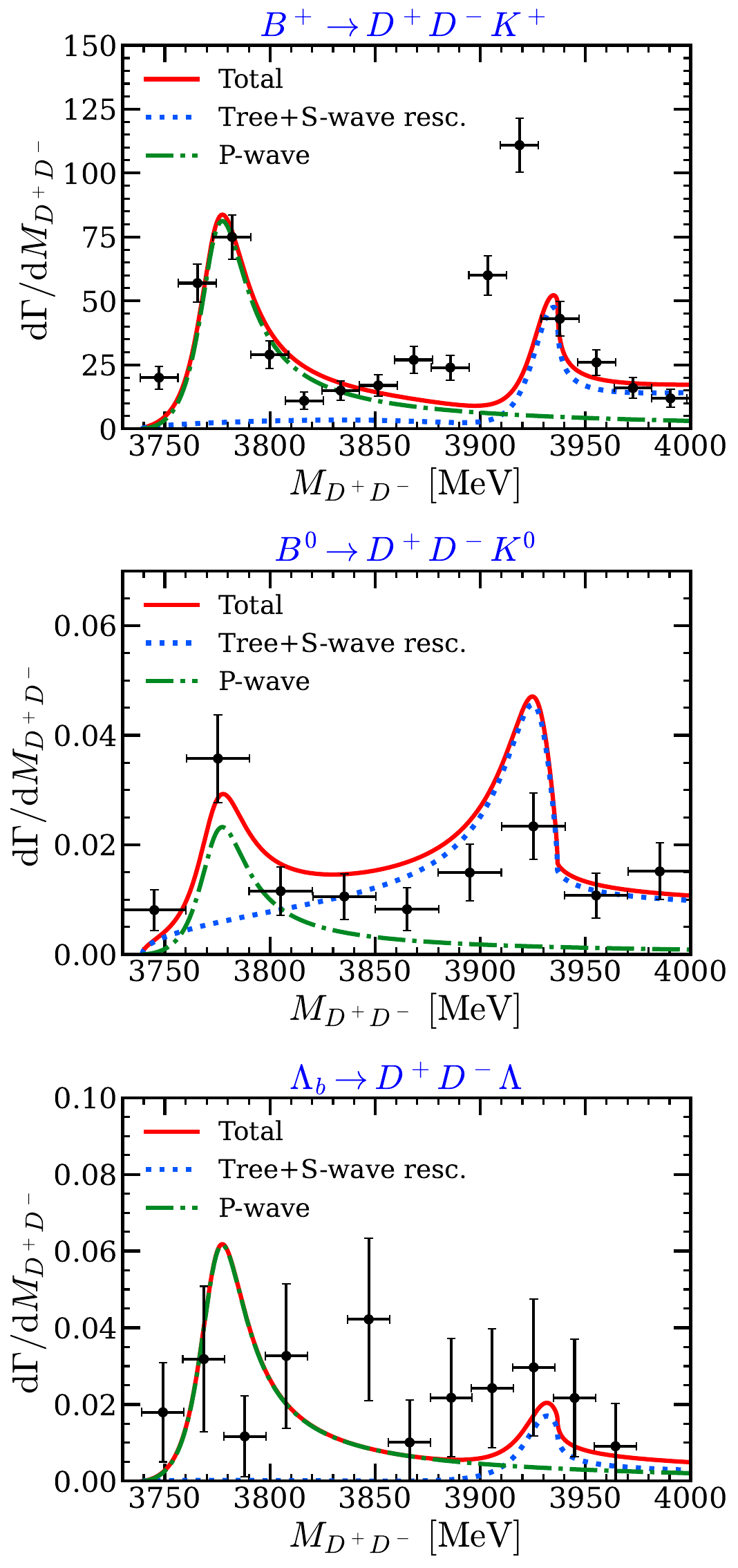}
  \caption{$D^+D^-$ invariant mass distribution, same as Fig.~\ref{Fig:comFit} but without $X(3700)$.}
  \label{Fig:kill_X3700}
\end{figure}

\begin{figure}[htbp]
  \centering
  \includegraphics[width=0.48\textwidth]{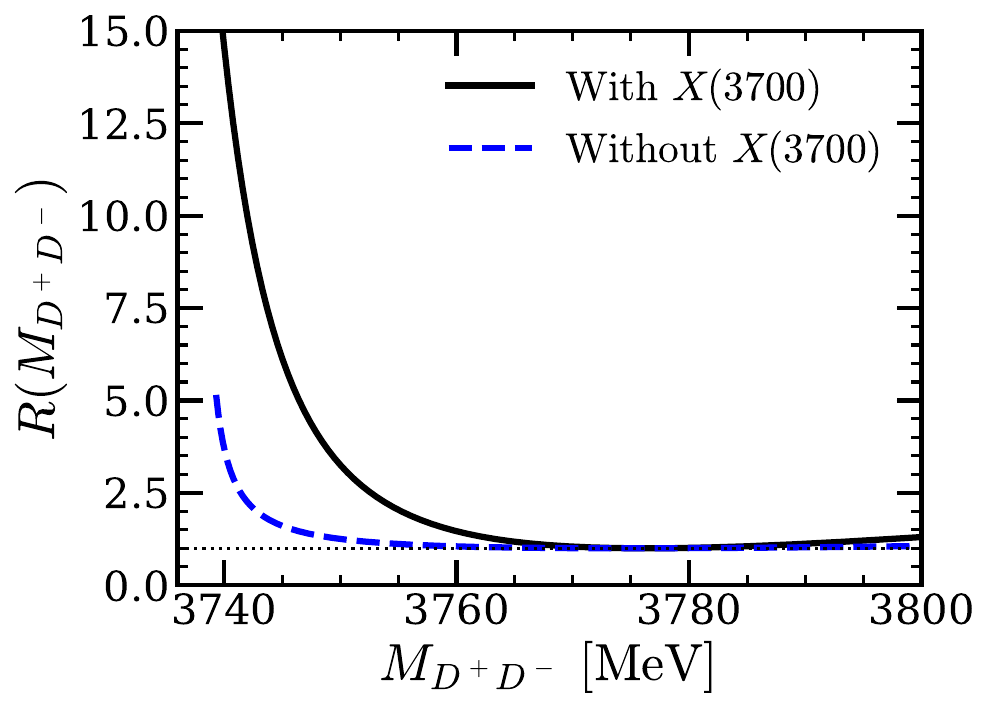}
  \caption{Comparison of the ratio of $D^+D^-$ invariant mass distributions with (solid line) and without (dashed line) the $X(3700)$ state for the $B^+$ and $\Lambda_b$ decay processes.}
  \label{Fig:ratio_kill_X3700}
\end{figure}

Next, we investigate the scenario where the $X(3700)$ state is omitted.
To obtain the amplitudes without $X(3700)$, we set $V_{ij}=0\,(i,j=1,2)$, and adjust $a_{D_s \bar{D}_s}=-0.86$ in Eq.~\eqref{Eq:ScenarioIII} of Scenario III to keep the $X(3930)$ at the right position.
The fitted results are shown in Figs.~\ref{Fig:kill_X3700} and \ref{Fig:ratio_kill_X3700}.
From Fig.~\ref{Fig:kill_X3700}, we can appreciate changes in the line-shapes of the mass distributions with respect to Fig.~\ref{Fig:comFit} which contains the $X(3700)$.
The most relevant information, however, is found in Fig.~\ref{Fig:ratio_kill_X3700} where,
as can be clearly seen, the ratio of invariant mass distributions undergoes a significant change when the $X(3700)$ is switched off.
The ratio in the absence of the $X(3700)$ goes up to about $5$ at threshold, but it is about $1.5$ at $M_{D^+D^-}=3745\, {\rm MeV}$, while when considering the $X(3700)$ state, it is around $7$.


\begin{figure}[htbp]
  \centering
  \includegraphics[width=0.48\textwidth]{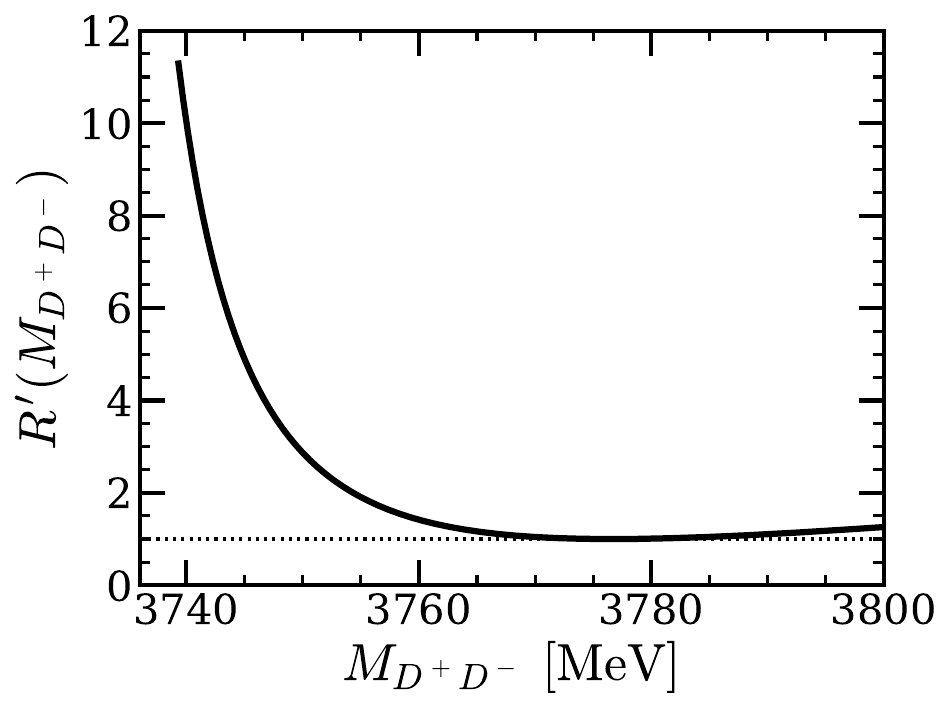}
  \caption{Ratio $R'(M_{D^+ D^-})$ of Eqs.~\eqref{eq:ratioRprime} and \eqref{eq:ratioRprime_tilde}.}
  \label{Fig:kill_X3700-R}
\end{figure}

The size of the ratio $R(M_{D^+D^-})$ in Fig.~\ref{Fig:ratio_kill_X3700} in the absence of the $X(3700)$ state is a bit disturbing, because it might suggest the presence of a state below threshold when there is none.
For this reason, we go back to the $B^+ \to D^+ D^- K^+$ reaction and show the results with and without the $X(3700)$ state.
We already showed in Figs.~\ref{Fig:x} and \ref{Fig:comFit} the results for the $D^+ D^-$ mass distribution of the $B^+ \to D^+ D^- K^+$ reaction with the three scenarios,
and we could see that the differences are small close to threshold, particularly between the scenarios II and III.
We also showed the $D^+D^-$ mass distribution for scenario III in Figs.~\ref{Fig:comFit} and \ref{Fig:kill_X3700} with the $X(3700)$ and without it.
To make the differences more explicit, we define the ratio for the $B^+\to D^+D^-K^+$ reaction as 
\begin{equation}\label{eq:ratioRprime}
R'(M_{D^+D^-}) = \frac{\tilde{R}'(M_{D^+D^-})}{\tilde{R}'(M_{D^+D^-} = 3770 \text{ MeV})},
\end{equation}
with
\begin{equation}\label{eq:ratioRprime_tilde}
\tilde{R}'(M_{D^+D^-}) = \frac{{\dd\Gamma}/{\dd M_{D^+D^-}}\left(\text{with } X(3700)\right)}{{\dd \Gamma}/{\dd M_{D^+D^-}}\left(\text{without } X(3700)\right)}.
\end{equation}
We plot this ratio $R'(M_{D^+D^-})$ in Fig.~\ref{Fig:kill_X3700-R}. 
We can see that the ratio grows steadily from $1$ at $3770\, {\rm MeV}$ (by construction) to values around $11$ at threshold, indicating a strong enhancement effect from the presence of the $X(3700)$ state.
Experimentally the test can be done as follows.
Take $\frac{\dd\Gamma}{\dd M_{D^+D^-}} (\text{without } X(3700))$ from our calculated result (Fig.~\ref{Fig:kill_X3700}) and divide the experimental $D^+D^-$ mass distribution by this theoretical mass distribution, both of them normalized to the same value at $M_{D^+D^-} = 3770 \text{ MeV}$.
The test can be also done using any theoretical model for the $B^+ \to D^+ D^- K^+$ reaction which reproduces fairly the $\psi(3770)$ peak and does not contain the $X(3700)$ state.

There is another issue that we would like to raise here about possible contributions of resonances in other channels than $D^+D^-$, or triangle mechanisms.
Actually, concerning this latter issue, it was shown in Ref.~\cite{Liu2020} that there is a triangle mechanism that enhances the $D^-K^+$ invariant mass around the energy of the $X_0(2900)$ and $X_1(2900)$ resonances.
Actually, it is easy to see that, with or without this mechanism, the consideration of these resonances does not influence our predictions at the threshold of the $D^+D^-$ mass distribution.
To prove this, we rely upon the Dalitz plot in Fig.~6 of the expermiental paper \cite{LHCb:2020pxc}.
There one can see, indeed, an enhanced contribution at $m^2(D^-K^+)\approx 8.5 \text{GeV}^2$, corresponding to the $X_0(2900)$ and $X_1(2900)$ resonances.
In the same figure, we can see that the region of the Dalitz plot for $m^2(D^-K^+)\approx8.5\ \text{GeV}^2$ has no overlap with the region of $m(D^+D^-)\simeq 3740\, {\rm MeV}$ ($m^2(D^+D^-)\simeq 14\ \text{GeV}^2$),
such that the presence of those resonances in the $D^-K^+$ mass distribution cannot affect the $D^+D^-$ mass distribution around threshold.

\begin{figure}[t]
    \centering
    \includegraphics[width=0.25\textwidth]{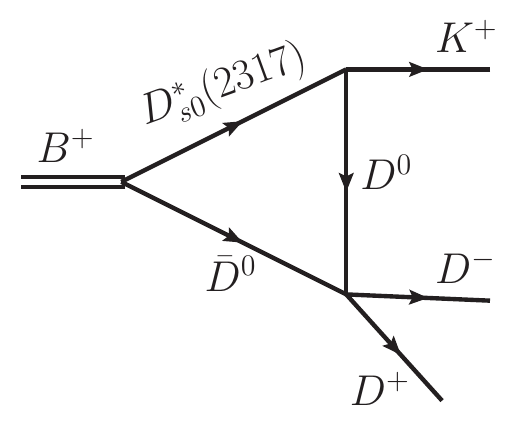}
    \caption{Triangle diagram leads to $B^+$ decay into $K^+D^+K^-$.}
    \label{fig:TS}
\end{figure}

It would be possible to find a triangle mechanism that enhances the $D^+D^-$ mass distribution at threshold.
We made a thorough search and found indeed a possible mechanism depicted in Fig.~\ref{fig:TS}.
Actually, the $B^+$ decays into $D^*_{s0}(2317)\bar{D}^0$ is reported in the PDG with a branching ratio of $7.9\times 10^{-4}$, not too small.
The $D_{s0}^*(2317)$ decays to $D^0 K^+$, and $D^0 \bar D^0$can produce $D^+D^-$, enhanced by the presence of the $X(3700)$ resonance below threshold.
However, there is no need to perform this calculation, because a triangle diagram can only produce relevant effects if it is close to a triangle singularity.
Yet, the diagram of Fig.~\ref{fig:TS} is very far away from this situation.
Indeed, to test how far it is, we apply Eq.~(18) of Ref.~\cite{Bayar2016}.
In the closest kinematics to a possible triangle singularity when the intermedicate particles are placed on-shell in a parallel situation,
we find that the $\bar D^0$ momentum from $B^+\to D^*_{s0}(2317)\bar D^0$ decay and from the $\bar D^0 D^0\to D^+D^-$ at the $D^+D^-$ threshold, 
which should be equal to have the triangle singularity, differ by more than $500 \, {\rm MeV}$, and in this most favorable kinematical case, 
the corresponding energy of the $B^+$ in Fig.~\ref{fig:TS} is $4243 \, {\rm MeV}$, far away from the nominal mass $5279 \, {\rm MeV}$ of the $B^+$, by more than $1000 \, {\rm MeV}$.
Conversely, if we choose the kinematical situation in which the energy of the $B^+$ would be the physical mass, the two $\bar D^0$ momenta differ by more than $1150\text{ MeV}$.

\section{Conclusions}
We have studied the $B^+ \to D^+ D^- K^+$, $B^0 \to D^+ D^- K^0$ and $\Lambda_b \to D^+ D^- \Lambda$  reactions, considering the final state interaction of the charmed meson pairs. 
This interaction gives rise to the $X(3930)$  that couples mostly to $D_s\bar{D_s}$ and another state $ X(3700)$ that couples mostly to $D\bar{D}$. 
The latter state appears theoretically in many calculations and in some lattice QCD results, and experimental evidence for it has also been claimed from the data of $\gamma \gamma \to D \bar D$~\cite{Deineka:2021aeu}.
Awaiting some evidence from a peak coming from decay of that state into pairs of light pseudoscalars~\cite{Xiao:2012iq}, one has to rely upon reactions producing $D \bar D$ pairs close to threshold. The  $B^+ \to D^+ D^- K^+$, $B_0 \to D^+ D^- K^0$ and $\Lambda_b \to D^+ D^- \Lambda$  reactions are in principle ideal to make this test, and this is why we study them here. 
Unfortunately, only 30 MeV above the $D^+ D^-$ threshold, the data show a large contribution from the $\psi(3770)$ vector meson which has a width of 27.3 MeV. 
Thus, the expected enhancement of the $D^+ D^-$ mass distribution close to threshold is not easy to see, and certainly is not visible in the limited statistics and resolution of the present data.
    
In spite of this, we have carried a study of the three reactions, looking in detail at the production mechanisms and the final state interaction of the $D \bar D$ and $D_s\bar{D_s}$ pairs, and we have seen that this interaction is relatively magnified in the  $B^+ \to D^+ D^- K^+$ reaction compared to that of the $\Lambda_b \to D^+ D^- \Lambda$  reaction. 
In view of this, we compare the $D^+ D^-$ mass distributions close to threshold for the two reactions and we observe that there is a clear enhancement very close to threshold in the first reaction which is not present in the latter one.
We have  quantified the enhancement by normalizing the two mass distributions at the peak of the $\psi(3770)$ and find that close to threshold the mass distribution for the $B^+ \to D^+ D^- K^+$ reaction is about an order of magnitude bigger than that of the $\Lambda_b \to D^+ D^- \Lambda$  reaction.  
The experimental test of this prediction demands precise data in the invariant mass range $M_{D^+ D^-} \in [3739,\,3750]$ MeV, where right now there is only one data point. 
This test should be feasible in next upgrades of the LHCb facility, and the observation of this predicted relative enhancement would provide great support to the existence of the  $D^+ D^- $ bound state. 
We also propose a different test by looking only to the $D^+ D^-$ mass distributions of the $B^+ \to D^+ D^- K^+$ reaction. 
In this case, the test should be done by dividing the experimental $D^+D^-$ mass distribution by our theoretical one constructed without the $X(3700)$ state, 
or any reasonable theoretical model that accounts for the $\psi(3770)$ excitation and does not contain the $X(3700)$ state,
where a large enhancement of this ratio values around $10$ at threshold are expected, when the two mass distributions are normalized to the same value at $M_{D^+D^-}=3770 \, {\rm MeV}$.
The present work should give an incentive to perform these measurements in the coming LHCb upgrades. 

\begin{table*}[t]
\centering
\caption{Pole positions of $X(3700)$ and $X(3930)$ states in the complex energy plane for three different scenarios.}
\label{tab:2and3channels}
\begin{tabular}{l|cc|cc|cc}
\hline\hline 
 & \multicolumn{2}{c}{Scenario I} 
& \multicolumn{2}{|c}{Scenario II} 
& \multicolumn{2}{|c}{Scenario III}\\[1mm]
 \cline{2-3}  \cline{4-5} \cline{6-7}
 & ~~$M$ [MeV] ~&~ $\Gamma$ [MeV]~~ &  ~~$M$ [MeV]~ &~ $\Gamma$ [MeV] ~~&  ~~$M$ [MeV] ~&~ $\Gamma$ [MeV]\\[1mm]
\hline 
Pole I [$X(3700)$]& 3725.4 & $\cdots$ & 3723.1  & 21.6 & 3724.1 & 26.2 \\[1.5mm]
Pole II [$X(3930)$] & 3930.4 & 13.7 & 3903.6  & 82.3 & 3931.4  & 44.6\\
\hline\hline 
\end{tabular}
\end{table*}

\begin{figure*}[htbp]
  \includegraphics[width=0.75\textwidth]{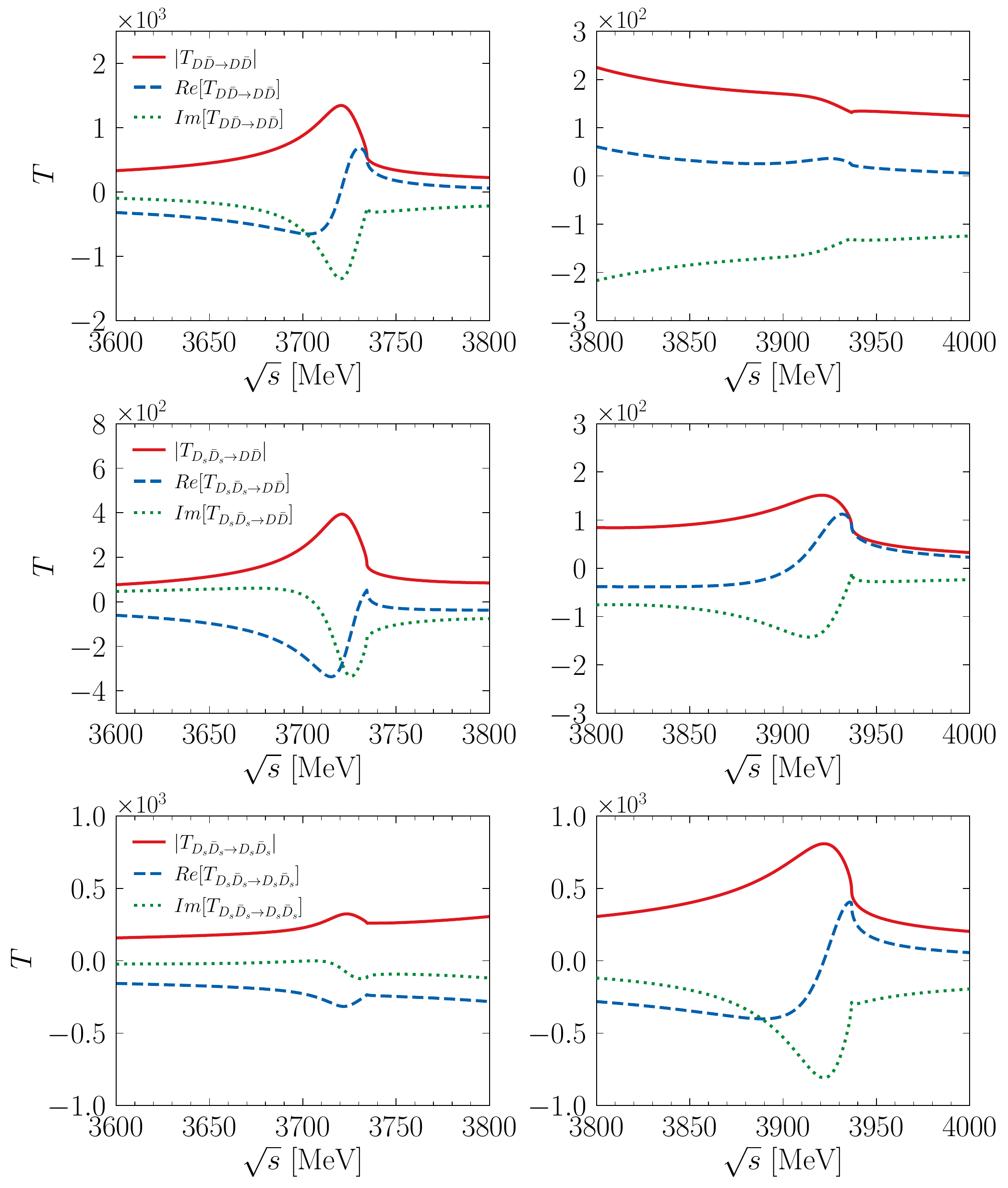}
  \caption{Two-body scattering amplitudes as a function of $\sqrt{s}$ obtained for Scenario III}. Left column: $\sqrt{s}\in [3600,\, 3800]$ MeV; right column: $\sqrt{s}\in [3800,\, 4000]$ MeV.
  \label{fig:amp}
\end{figure*}

\acknowledgements 
This work is partly supported by Shandong Provincial Natural Science Fund for Excellent Young Scientists Fund Program (Overseas) with project no. 2025HWYQ-015,  and  by Qilu Youth Scholars Program of Shandong University.
This work is partly supported by the National Natural Science Foundation of China (NSFC) under Grants No. 12575081 and No. 12365019, and by the Natural Science Foundation of Guangxi province under Grant No. 2023JJA110076. This work is partly supported the National Key R\&D Program of China (Grant No. 2024YFE0105200).
This work is also partly supported by the Spanish Ministerio de Economia y Competitividad (MINECO) and European FEDER funds under Contracts No. FIS2017-84038-C2-1-PB, PID2020-112777GB-I00, and by Generalitat Valenciana under contract PROMETEO/2020/023. 
This project has received funding from the European Union Horizon 2020 research and innovation program under the program H2020-INFRAIA-2018-1, grant agreement No. 824093 of the STRONG-2020 project.

\appendix
\section{Three scenarios for the dynamical generation of the $X(3700)$ and $X(3930)$ states and the two-body scattering amplitudes}\label{APP}

We follow Ref.~\cite{Bayar:2022dqa} including also the $\eta\eta$ channel to the $D\bar{D}$ and $D_s\bar{D}_s$, as done in Ref.~\cite{Xiao:2012iq}. 
The transition matrix is given by   
\begin{equation}
\begin{pmatrix}
	\eta\eta \to \eta  \eta & \eta\eta \to D  \bar{D}  & \eta\eta \to D_s \bar{D}_s \\[1mm]
	D \bar{D} \to \eta  \eta & D \bar{D} \to D  \bar{D}  & D \bar{D} \to D_s \bar{D}_s   \\[1mm]
	D_s \bar{D}_s \to \eta \eta & D_s \bar{D}_s \to D \bar{D} &  D_s \bar{D}_s \to D_s \bar{D}_s   
\end{pmatrix}.
\end{equation}

Then, we have the $S$-wave interaction and $V_{ij}$ reads as follows  
\begin{widetext}
\begin{equation}
  \label{eq:Veta1}
\begin{aligned}
	V =
\begin{pmatrix}
0 & -v_\eta & 0  \\[2mm]
-v_\eta & -\frac{1}{4}\left(\frac{3}{M_\rho^2}+\frac{1}{M_\omega^2}+\frac{2}{M_{J / \psi}^2}\right)  \biggl(3s - 4M_{D}^2 \biggr) & -\frac{\sqrt{2}}{2} \frac{1}{M_{K^*}^2+M_{D_s}^2 - M_D^2 }  \biggl(3s - 2(M_{D}^2+M_{D_s}^2)  \biggr) \\[3.5mm]
0 & -\frac{\sqrt{2}}{2} \frac{1}{M_{K^*}^2+M_{D_s}^2 - M_D^2 }  \biggl(3s - 2(M_{D}^2+M_{D_s}^2)  \biggr)
& -\frac{1}{2}\left(\frac{1}{M_\phi^2}+\frac{1}{M_{J / \psi}^2}\right) \biggl(3s - 4M_{D_s}^2 \biggr) \\
\end{pmatrix},
\end{aligned}
\end{equation}
\end{widetext}
from where we see that, for the transitions involving $\eta \eta$ channel, we only have the nonzero transition between $\eta\eta$ and $D\bar{D}$.

The two-body scattering $t$-matrix is given by 
\begin{equation}
	t  = \left[ 1 - VG \right]^{-1} V,
\end{equation}
and we use $G$, the loop function of intermediate mesons in dimensional regularization~\cite{Gamermann:2006nm}.

Next, we present the pole positions of $X(3700)$ and $X(3930)$ with different regularization parameters in the $G$ loop functions,  corresponding to the following three scenarios.
\begin{itemize}
  \item Scenario I: Two coupled channels $D\bar{D}$ and $D_s\bar{D}_s$, with $v_\eta=0$ in Eq.~\eqref{eq:Veta1}, and the same regularization parameters in Ref.~\cite{Abreu:2025jqy},  
  \begin{equation}
  \label{Eq:ScenarioI}
    \mu=1500~\mathrm{MeV}, \, a_{D\bar{D}} = -0.8, \, a_{D_s\bar{D}_s}=-1.58;
  \end{equation}
  \item Scenario II: Three coupled channels $D\bar{D}$, $D_s\bar{D}_s$ and $\eta\eta$, with $v_\eta=200$, and the regularization parameters as   
  \begin{equation}
  \label{Eq:ScenarioII}
    \mu=1500~\mathrm{MeV}, \, a_{\eta\eta} =a_{D\bar{D}} = -0.8, \, a_{D_s\bar{D}_s}=-1.58;
  \end{equation}
  \item Scenario III: Same as Scenario II, but with slightly different $a_{D_s\bar{D}_s}$ adjusted to get the right position of the $X(3930)$, 
  \begin{equation} 
  \label{Eq:ScenarioIII}
    \mu=1500~\mathrm{MeV}, \, a_{\eta\eta} =a_{D\bar{D}} = -0.8, \, a_{D_s\bar{D}_s}=-1.4.
  \end{equation}
\end{itemize}

In Table~\ref{tab:2and3channels}, we tabulate the predictions of pole positions of $X(3700)$ and $X(3930)$ in the complex energy plane for the above three scenarios. 
The two-body scattering amplitudes corresponding to Scenario III are given in Fig.~\ref{fig:amp}.  
We prefer to employ these two-body scattering amplitudes in the calculations of $B^{+}$, $B^0$, and $\Lambda_b$ decays. 

\bibliographystyle{a}
\bibliography{ref}

\end{document}